\documentclass[fleqn,usenatbib]{mnras}

\usepackage{newtxtext,newtxmath}

\usepackage[T1]{fontenc}
\usepackage{natbib}

\DeclareRobustCommand{\VAN}[3]{#2}
\let\VANthebibliography\thebibliography
\def\thebibliography{\DeclareRobustCommand{\VAN}[3]{##3}\VANthebibliography}

\usepackage{graphicx}	
\usepackage{amsmath}	
\usepackage{multicol}
\usepackage{hyperref}

\title[Eccentric ZLK effects]{Eccentric von Zeipel--Lidov--Kozai effects under mildly hierarchical triple systems: Influence of Brown corrections upon orbit flipping}

\author[H. Gao and H. Lei]{
Hao Gao$^{1,2}$,
Hanlun Lei$^{1,2}$\thanks{E-mail: \href{mailto:leihl@nju.edu.cn}{leihl@nju.edu.cn}}
\\
$^{1}$ School of Astronomy and Space Science, Nanjing University, Nanjing 210023, China\\
$^{2}$ Key Laboratory of Modern Astronomy and Astrophysics in Ministry of Education, Nanjing University, Nanjing 210023, China
}

\date{Accepted XXX. Received YYY; in original form ZZZ}

\pubyear{\the\year{}}

\begin{document}
\label{firstpage}
\pagerange{\pageref{firstpage}--\pageref{lastpage}}
\maketitle

\begin{abstract}
Mildly hierarchical three-body systems are widespread in the Universe, exemplified by planets in stellar binaries and stars in black-hole binaries. In such systems, Brown Hamiltonian corrections play a crucial role in governing the long-term dynamical evolution. In this work, we extend Brown corrections to include octupole-order coupling terms, thereby formulating a more accurate dynamical model for predicting long-term dynamical behaviors. The utilization of the gauge freedom in canonical transformation shows that the quadrupole-octupole coupling term vanishes and the octupole-octupole coupling term is axisymmetric. Under triple systems with different levels of hierarchies, we systematically investigate the impact of Brown corrections on orbital flipping induced by the eccentric von Zeipel–Lidov–Kozai (ZLK) mechanism. Our analysis reveals that, as the hierarchy of triple systems becomes lower, the asymmetry in the flipping regions becomes more significant. The asymmetric structures are examined in detail using Poincar\'e sections and perturbative techniques, showing that Brown corrections are the key factor responsible for breaking the symmetry of flipping regions. Finally, we extend the classical pendulum approximation to our refined model and demonstrate that its analytical predictions agree remarkably well with those derived from perturbative methods, particularly in the high-eccentricity regime.
\end{abstract}




\section{Introduction}
\label{sec:0}

Due to long-term perturbations from a distant tertiary companion, the inclination and orbital eccentricity of the inner binary undergo coupled oscillations. When the mutual inclination lies between approximately $39.2^\circ$ and $140.8^\circ$, these perturbations can trigger large-amplitude variations in both inclination and eccentricity. Such a phenomenon is known as the von Zeipel–Lidov–Kozai (ZLK) effect \citep{ito2019}. This dynamical mechanism was originally discovered by \citet{von1910} and later independently rediscovered by \citet{Kozai1962} and \citet{Lidov1962}. Please refer to \citet{naoz2016} and \citet{shevchenko2016lidov} for review of the ZLK effect and its applications in a wide range of astrophysical systems.

Classical studies of ZLK oscillations assume that the orbit of the third body is nearly circular \citep{Kozai1962}. Under the test-particle approximation, the secular evolution of the outer orbit is neglected, and the gravitational potential from the tertiary companion is typically truncated at the quadrupole order with respect to the semimajor axis ratio between the inner and outer binaries. This axisymmetric, quadrupole-level Hamiltonian is independent of the longitude of the ascending node, resulting in the conservation of the $z$-component of the orbital angular momentum. As a consequence, the orbit of the inner binary remains confined to either prograde or retrograde configurations, thereby prohibiting orbital flips. Even when the third body is in an elliptical orbit, this conclusion remains unchanged. Despite this simplification, the quadrupole approximation has been widely employed to explain dynamical phenomena across a broad range of astrophysical systems--including artificial satellites, Kuiper Belt objects, exoplanetary systems, stellar triples, and systems involving supermassive black holes \citep{naoz2016,shevchenko2016lidov}.

When the assumption of a circular orbit for the perturbing body is relaxed, the long-term evolution of the system exhibits qualitatively different behavior. In this case, the third-body disturbing function needs to be expanded up to the octupole order in the semimajor axis ratio, resulting in a Hamiltonian that explicitly depends on the longitude of the ascending node. As a consequence, the vertical component of angular momentum, $H$, is no longer conserved, and the classical ZLK oscillations become modulated over longer timescales \citep{katz2011long}. Over these extended timescales, the inner binary's orbit can undergo periodic flips between prograde and retrograde configurations. At the moment of orbital flipping, the eccentricity of the inner binary can be driven to near-unity values. This behavior was termed the `eccentric Kozai mechanism' by \citet{Lithwick&Naoz2011}, and now it is widely referred to as the eccentric von Zeipel–Lidov–Kozai (ZLK) effect \citep{ito2019}. Within the framework of the eccentric ZLK mechanism, triple systems can exhibit chaotic dynamics, particularly in high-inclination regimes \citep{li2014chaos}. \citet{sidorenko2018eccentric} interpreted this effect as a resonance phenomenon, and from this resonant perspective, \citet{lei2022dynamical} constructed a global map of octupole-order resonances in the eccentricity–inclination space. Moreover, \citet{lei2022} performed a systematical investigation of orbital flips through various approaches, including Poincar\'e sections, periodic orbits and their invariant manifolds, and perturbative techniques. Their results showed that flipping orbits driven by the eccentric ZLK effect correspond to quasi-periodic trajectories that librate around polar periodic orbits. More recently, \citet{huang2024dynamical} reported bifurcations of periodic orbits associated with ZLK oscillations. These bifurcations arise from ZLK secondary resonances, which have been systematically studied via perturbation methods \citep{zhao2024zeipel}, showing that the ZLK secondary resonances can enhance the excitation of eccentricity.

In the study of long-term evolution under hierarchical systems, averaging techniques are commonly employed--namely, single averaging (SA), which involves averaging over the inner orbital period, and double averaging (DA), which averages over both the inner and outer orbital periods. In strongly hierarchical configurations, the timescale of ZLK oscillations is typically much longer than the orbital periods of both the inner and outer orbits, allowing the secular approximation to remain valid for analyzing long-term dynamics. However, as the hierarchical structure becomes less pronounced, the ZLK timescale shortens, leading to a breakdown in the separation of timescales between orbital motion and secular evolution. In such mildly hierarchical systems, second-order perturbations due to the evection terms in the single-averaged disturbing function become important in shaping long-term dynamical behavior. This issue is well known in the context of the lunar problem \citep{cuk2004secular,tremaine2023}. As discussed in \citet{tremaine2023}, this phenomenon was systematically studied by \citet{brown1936a,brown1936b,brown1936c}, who introduced a nonlinear correction term, known as the quadrupole–quadrupole coupling term, to the long-term Hamiltonian. In recognition of his foundational work, this extended Hamiltonian is now referred to as Brown’s Hamiltonian \citep{tremaine2023}.

In the literature, multiple formulations of Brown’s Hamiltonian have been developed \citep{soderhjelm1975three, krymolowski1999studies, cuk2004secular, breiter2015secular, luo2016double,will2021higher, lei2018modified}. It is demonstrated by \citet{tremaine2023} that these different formulations are related by a gauge freedom inherent in canonical transformations. Recently, \citet{conway2024higher} extended the Brown's model developed in \citet{will2021higher} to second order in the fundamental quadrupolar perturbation parameter, and to dotriacontapole order in the
semimajor axis ratio between the inner and outer orbits using a two-timescale method. Regarding low-hierarchy triple systems, \citet{lei2025Extensions} developed a high-precision dynamical model, incorporating the nonlinear effects arising from both the inner and outer binaries. This model is referred to as the extended Brown Hamiltonian model, where the Hamiltonian is expressed in an elegant and closed form with respect to the eccentricities of both the inner and outer orbits.

Within the Brown Hamiltonian framework, dynamical features including the maximum eccentricity, critical inclination, and fixed points have been analytically explored by \citet{grishin2018quasi} and \citet{grishin2024irregulara}. In the high-eccentricity regime, Brown’s Hamiltonian induces azimuthal precession of the eccentricity vector, and the resulting ZLK dynamics can be effectively approximated using a simple pendulum model \citep{klein2024hierarchicalb}. Moreover, modified ZLK oscillations under Brown’s Hamiltonian have found important applications in various astrophysical contexts. These include the long-term evolution of irregular satellites in the Solar System \citep{grishin2024irregularb}, the formation of binary black hole mergers \citep{su2025possible}, and the progenitor scenarios for Type Ia supernovae \citep{rajamuthukumar2023triple}.

 For mildly hierarchical three-body systems, when the semimajor axis ratio is not particularly small, it becomes necessary to extend Brown’s Hamiltonian to include higher-order coupling terms \citep{lei2018modified}. Different levels of correction can significantly affect the eccentric ZLK effects. To this end, we formulate a high-order Hamiltonian model up to the fifth order in the semimajor axis ratio, incorporating Brown corrections up to the octupole-order coupling level. This high-precision dynamical model provides a robust framework for investigating the long-term evolution of the inner binary. Within this framework, we examine how Brown corrections influence orbital flips driven by eccentric ZLK effects under different levels of hierarchies. Specifically, we explore the flipping regions using Poincar\'e sections and perturbative methods. Our results indicate that Brown corrections are responsible for breaking the symmetry of the flipping regions. 
 
It should be mentioned that the term `hierarchical ordering' in this work specifically denotes the disparity between the ZLK oscillation timescale and the Keplerian orbital timescale, where their ratio depends on both the semi-major axis ratio and mass ratio. Consequently, modifying either the masses while holding the semi-major axis constant or altering the semi-major axis with fixed masses changes the hierarchical level. This study employs the former scenario to probe the influence of the third body's mass.

This remaining part of this work is organized as follows. In Section \ref{sec:2}, the disturbing function of hierarchical triple systems is briefly introduced. In Section \ref{sec:3}, the long-term Hamiltonian up to the fifth order in semimajor axis ratio is formulated with consideration of Brown corrections up to the octupole-octupole coupling term. Orbit flipping caused by the eccentric ZLK effects is discussed in Section \ref{sec:4} under triple systems with different levels of hierarchies. In Section \ref{sec:5}, pendulum approximation is applied to our formulated Hamiltonian model. At last, conclusions are summarized in Section \ref{sec:6}.

\section{Disturbing function}
\label{sec:2}

In this work, we consider a restricted hierarchical three-body system, which includes an inner binary consisting of a central body with mass $m_{0}$ and a test particle with mass $m_{1}$, together with a distant perturbing body with mass $m_{2}$. In the test-particle limit, the influence of $m_1$ on the motion of $m_{2}$ relative to $m_0$ is negligible. As a result, the perturbing body $m_{2}$ moves around the central body $m_{0}$ on a fixed Keplerian orbit. 

For convenience of description, we define an $m_0$-centered non-rotating coordinate system where the orbital plane of $m_{2}$ is the fundamental $x$--$y$ plane with the $x$-axis aligned with the perturber's eccentricity vector, and the $z$-axis is parallel to the vector of orbital angular momentum. In this coordinate system, the classical orbital elements are used to describe the motion of $m_1$ and $m_2$, including the semi-major axis $a$, eccentricity $e$, orbital inclination $i$, longitude of the ascending node $\Omega$, argument of pericenter $\omega$, and mean anomaly $M$. Without otherwise specified, the variables with subscript 1 are used for the inner test particle $m_{1}$ and the ones with subscript 2 are for the perturbing body $m_{2}$. In order to formulate the Hamiltonian model, it is necessary to introduce the following set of Delaunay variables \citep{morbidelli2002modern}:
\begin{equation}\label{eq:2}
    \begin{aligned}
    &l={M_1},\quad g={\omega_1},\quad h={\Omega_1},\\
    &L=\sqrt{\mu{a_1}},\quad G=L\sqrt{{1-e_1^2}},\quad H=G \cos{i_1},     
    \end{aligned}
\end{equation}
where $\mu = {\cal G} m_0$ is the gravitational parameter of the central object. Due to the hierarchical configuration, it holds the relation ${a_2} \gg {a_1}$, showing that the semimajor axis ratio $\alpha = a_1/a_2$ is a small parameter. As a result, the disturbing function can be expanded in a power series of $\alpha$ by taking advantage of Legendre polynomial as follows \citep{harrington1969stellar}:
\begin{equation}
    {\cal R} = \frac{{\cal G}{m_2}}{a_2} \sum\limits_{n = 2}^\infty  {{{\left( {\frac{{{a_1}}}{{{a_2}}}} \right)}^n}{{\left( {\frac{{{r_1}}}{{{a_1}}}} \right)}^n}} {\left( {\frac{{{a_2}}}{{{r_2}}}} \right)^{n + 1}}{P_n}\left( {\cos \psi } \right),
        \label{eq:4}
\end{equation}
where ${\cal G}$ is the universal gravitational constant, $r_1$ and $r_2$ are the distances of $m_1$ and $m_2$ relative to the central object $m_0$, $\psi$ is the relative angle and its cosine function can be expressed as \citep{lei2018modified}
\begin{equation}
    \begin{aligned}
        \cos{\psi} &= \frac{1}{{1 - {e_1}\cos{E_1}}}\left[ {{A_1}\cos{E_1}\cos {f_2} + {A_2}\cos{E_1}\sin{f_2}} \right.\\
        &+\left. { {B_1}\sin{E_1}\cos{f_2} + {B_2}\sin{E_1}\sin{f_2} + {C_1}\cos{f_2} + {C_2}\sin{f_2}} \right]
            \label{eq:6}
    \end{aligned}
\end{equation}
with ${E_1}$ and ${f_2}$ as the eccentric anomaly of the test particle and the true anomaly of the perturbing body, respectively. Please refer to \citet{lei2018modified} for the detailed expressions of $A_i$, $B_i$ and $C_i$ with $i=1,2,3$.

\begin{figure*}
	\includegraphics[width=5.05cm]{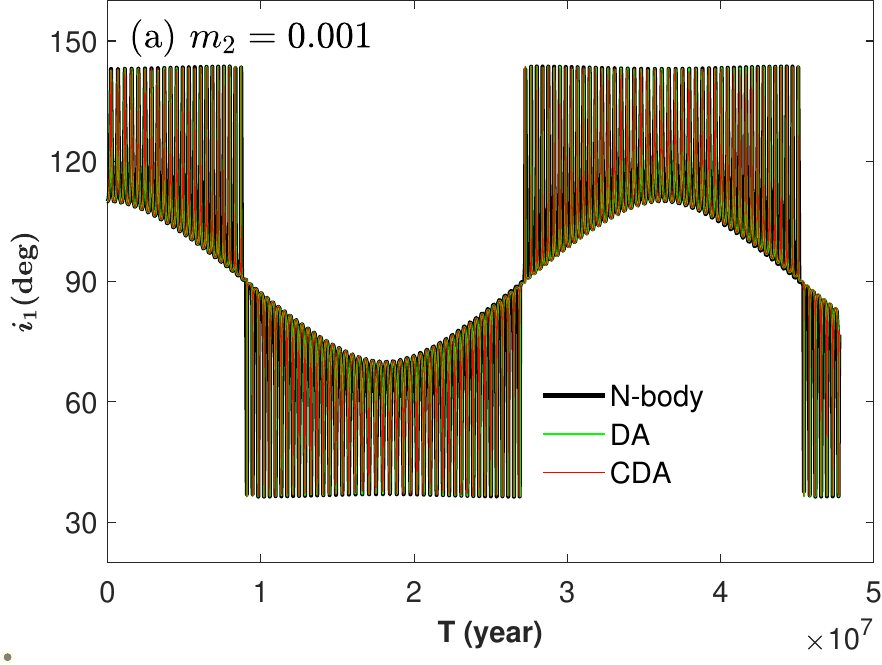}
    \includegraphics[width=5cm]{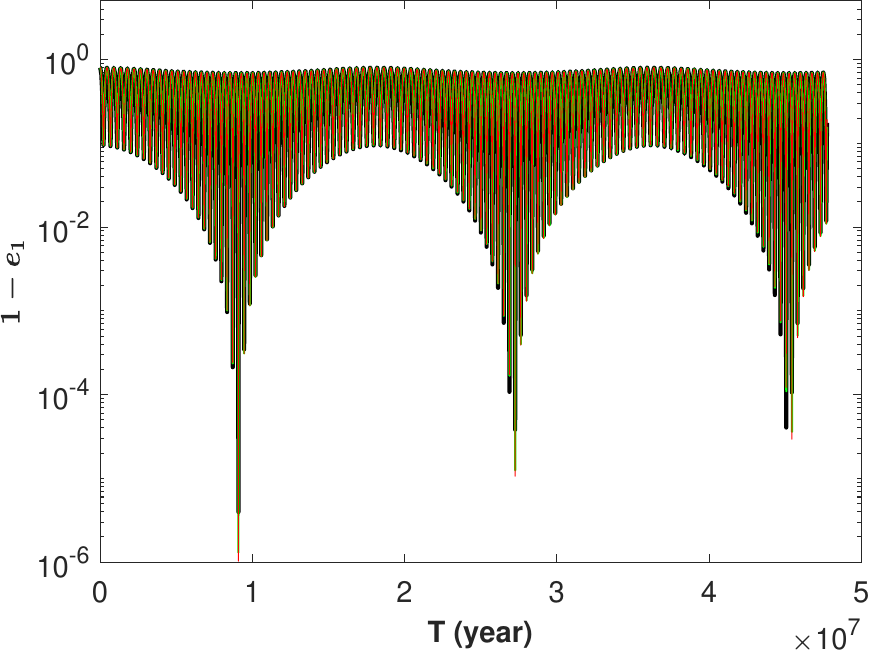}
    \includegraphics[width=5.15cm]{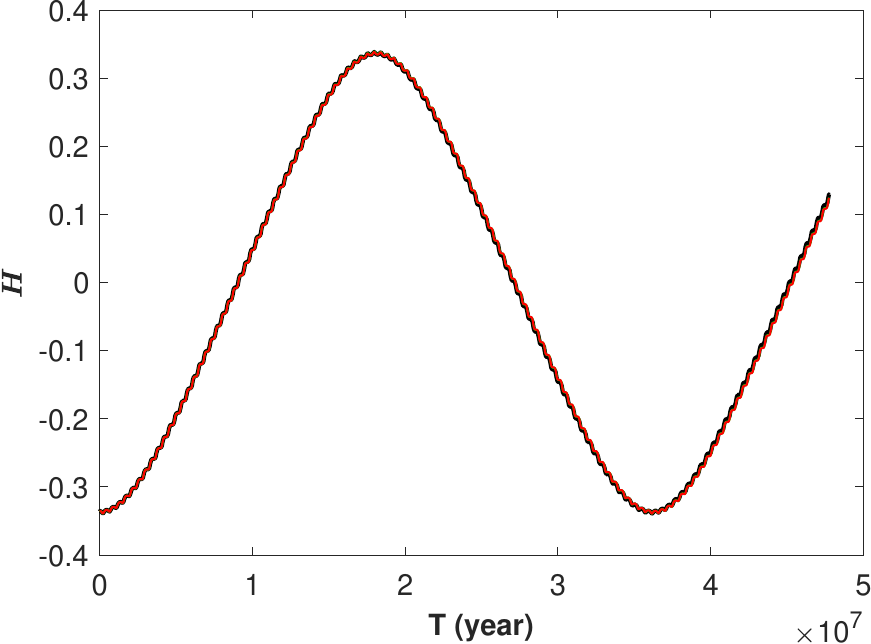}
        \\ \vspace{0.2cm}
        \includegraphics[width=5.05cm]{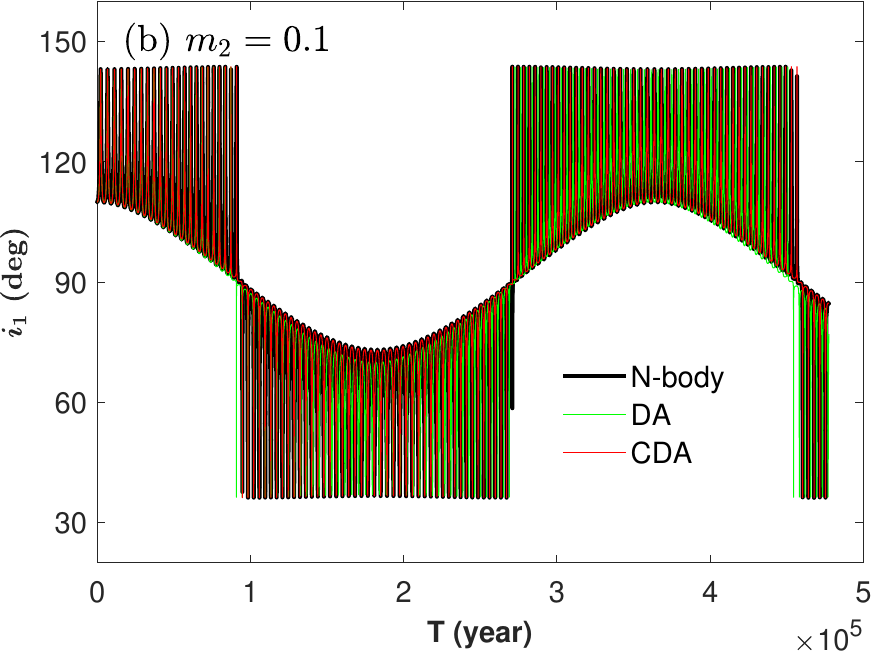}
        \includegraphics[width=5cm]{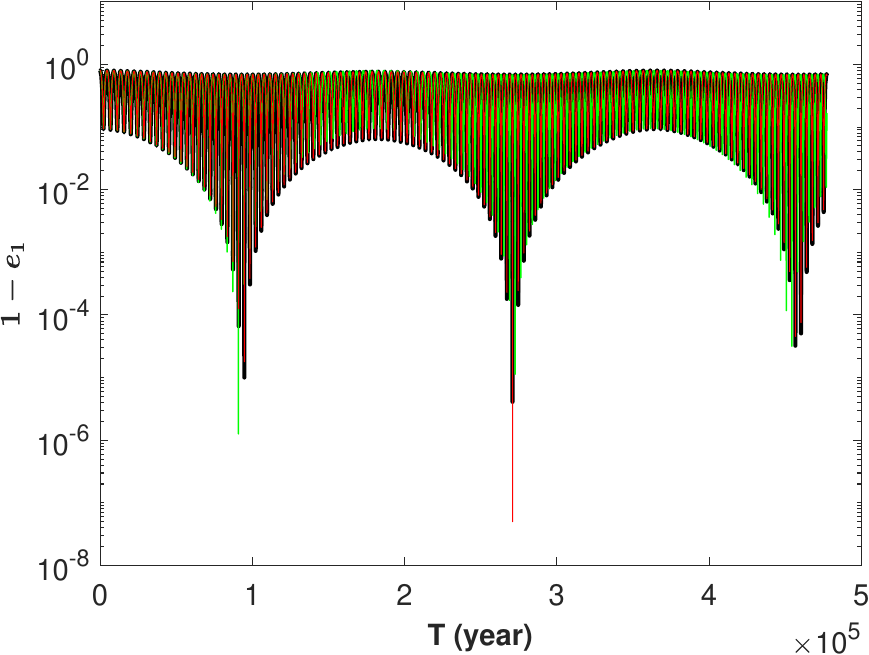}
        \includegraphics[width=5.15cm]{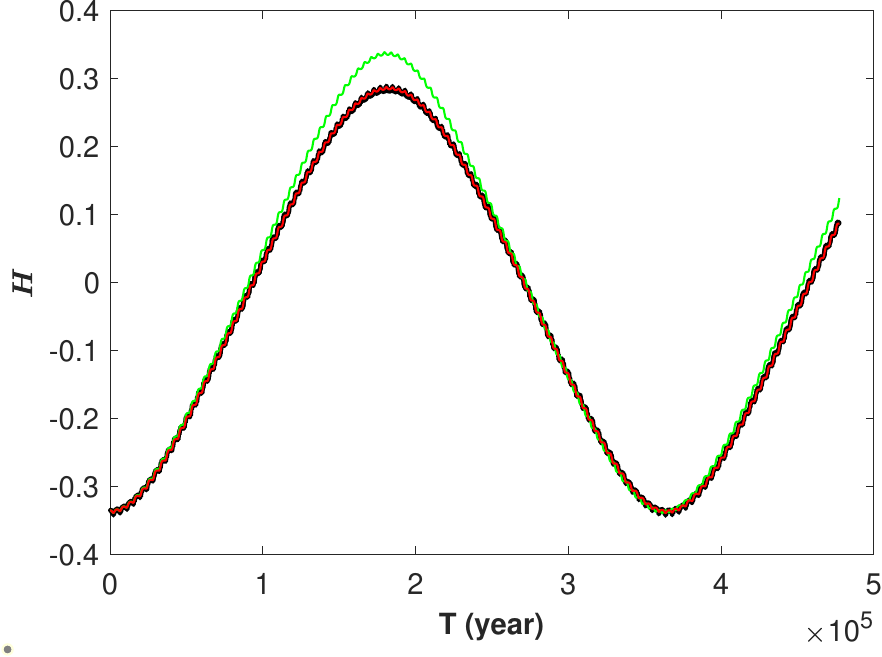}
        \\ \vspace{0.2cm}
        \includegraphics[width=5.05cm]{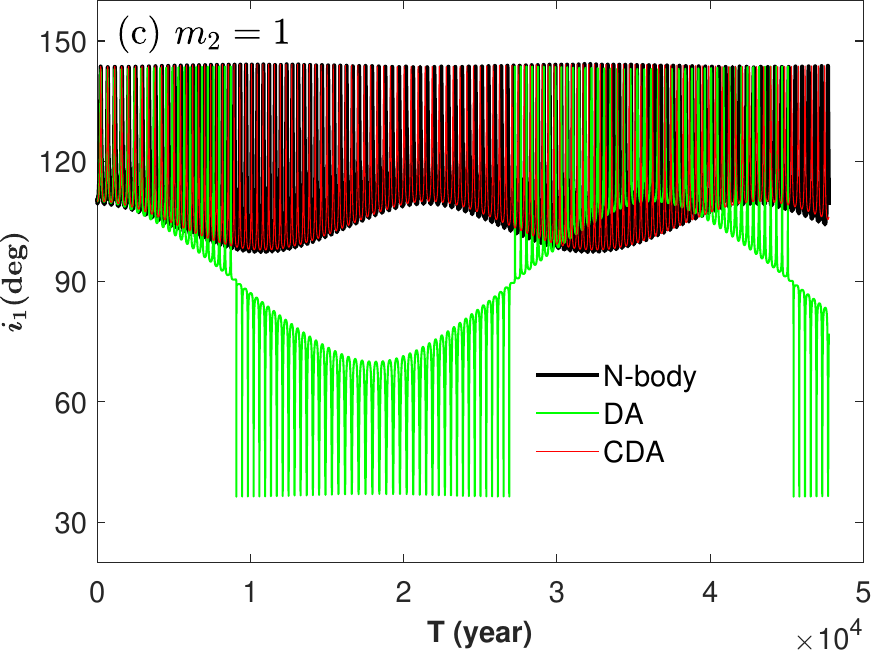}
        \includegraphics[width=5cm]{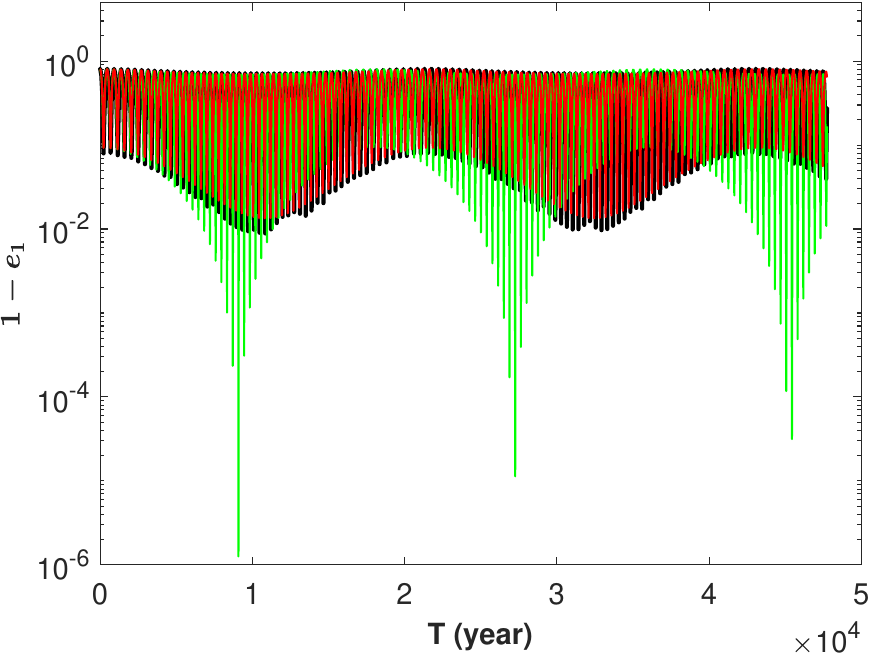}
        \includegraphics[width=5.15cm]{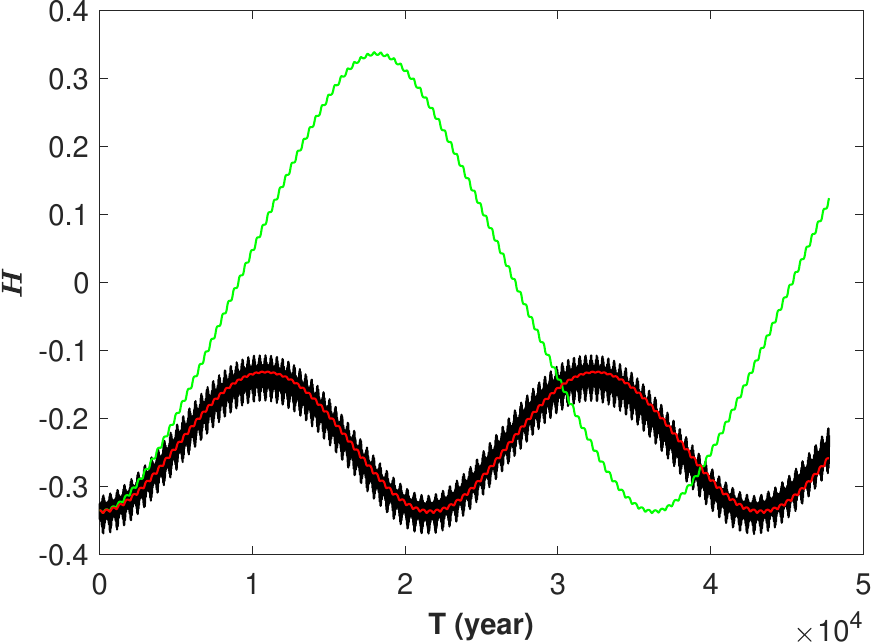}
    \caption{The temporal evolution of the test particle's inclination $i_1$ (\textit{left-column panels}), eccentricity $e_1$ (\textit{middle-column panels}) and $z$-component of the orbital angular momentum $H$ (\textit{right-column panels}). The mass of the central body is taken as $m_0=1.0\,m_\odot$. The initial orbital elements of the perturber are $a_2=10\text{AU}$ and $e_2=0.2$ (the remaining elements are assumed at zero). The mass $m_2$ is marked in the left-column panels in unit of solar mass. The initial orbital elements of the test particle are taken as $a_1=1.0 \text{AU},e_1=0.2,i_1=110^\circ,\Omega_1=180^\circ,\omega_1=0^\circ$, same as the ones adopted in \citet{luo2016double}. Green curves represent the results under the classical double-averaged models up to the fifth order in $\alpha$ without Brown correction (denoted by `DA'), red curves correspond to the results under the long-term dynamical model up to the fifth order in $\alpha$ with inclusion of Brown corrections (denoted by `CDA'), and black curves denote the $N$-body simulation results.}
    \label{fig:1}
\end{figure*}

\begin{figure}
    \includegraphics[width=8cm]{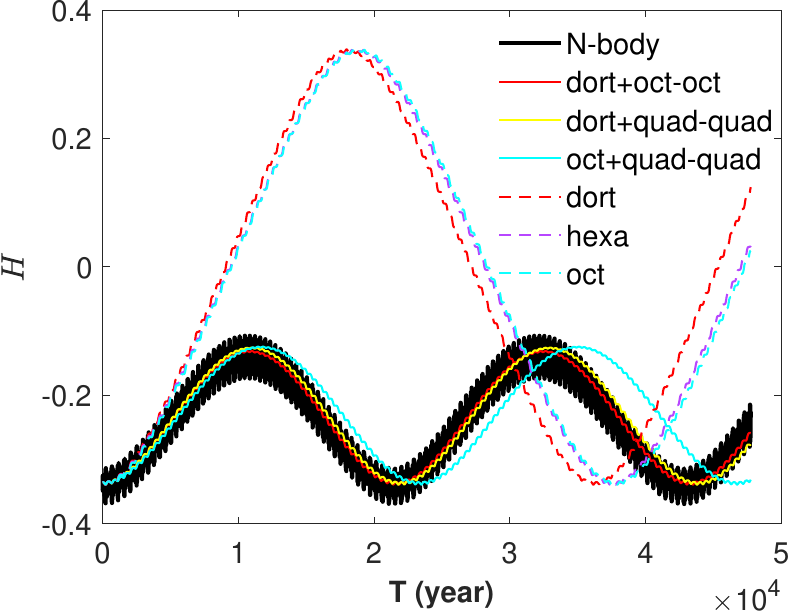}
    \caption{Comparison of models truncated at different orders. The mass of third body is taken as $m_2=1.0\,m_\odot$, while other parameters are selected identical to those adopted in Figure \ref{fig:1}. The solid red curve represents truncation to dotriacontapole and octupole-octupole coupling terms, the light blue curve corresponds to truncation to octupole and quadrupole-quadrupole coupling terms, and the yellow curve shows truncation to dotriacontapole and quadrupole-quadrupole coupling terms. Dashed curves in red, purple, and light blue depict models truncated exclusively to the dotriacontapole, hexadecapole, and octupole orders respectively, without correction terms.}
    \label{fig:9}
\end{figure}

\begin{figure*}
    \includegraphics[width=5.6cm]{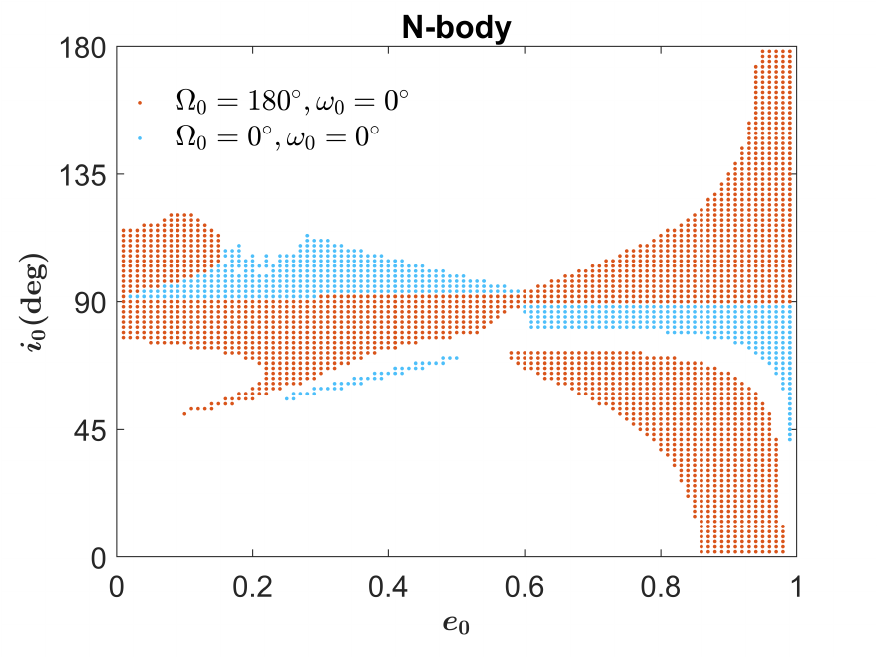}
    \includegraphics[width=5.6cm]{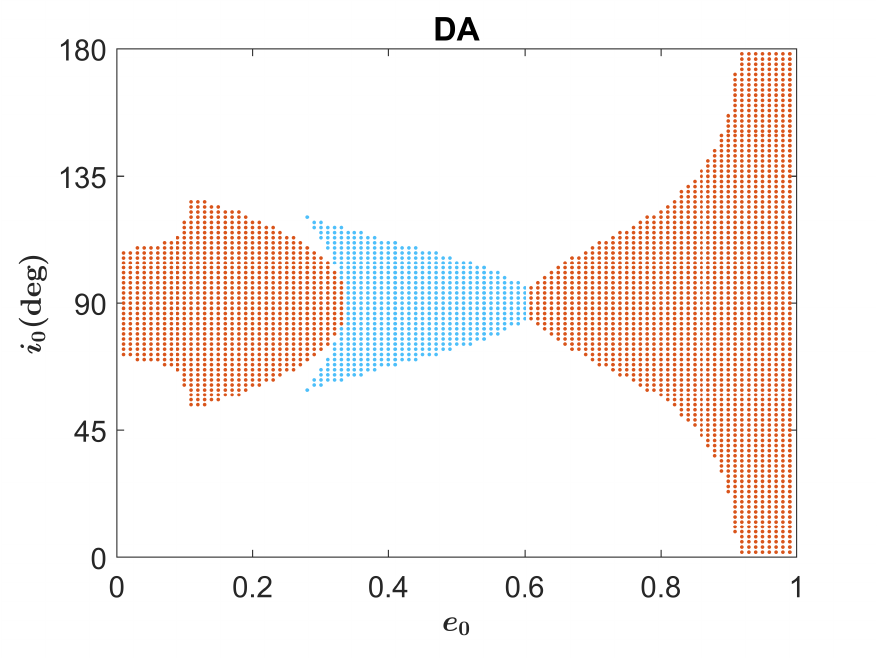}
    \includegraphics[width=5.6cm]{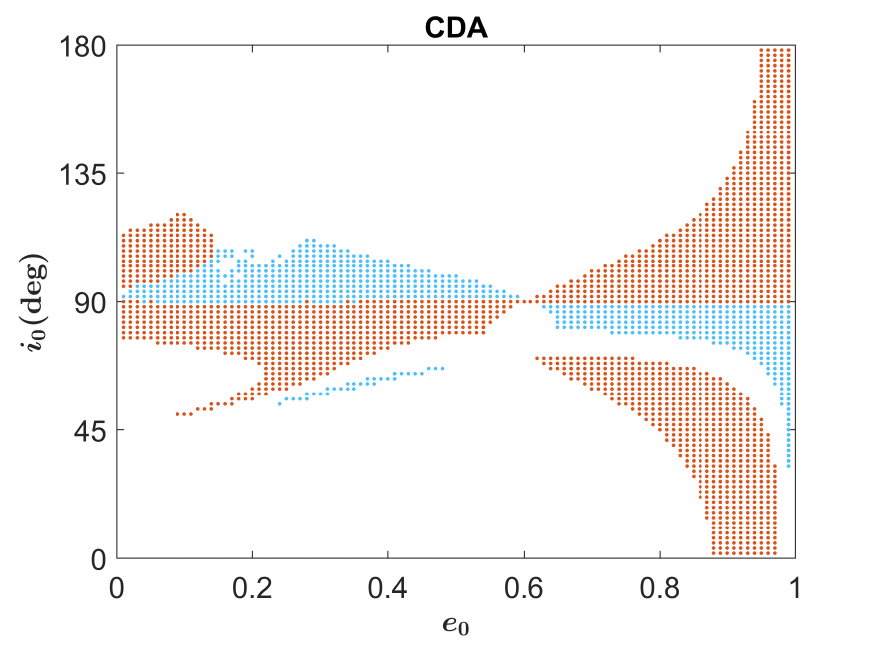}
    \\
    \includegraphics[width=5.6cm]{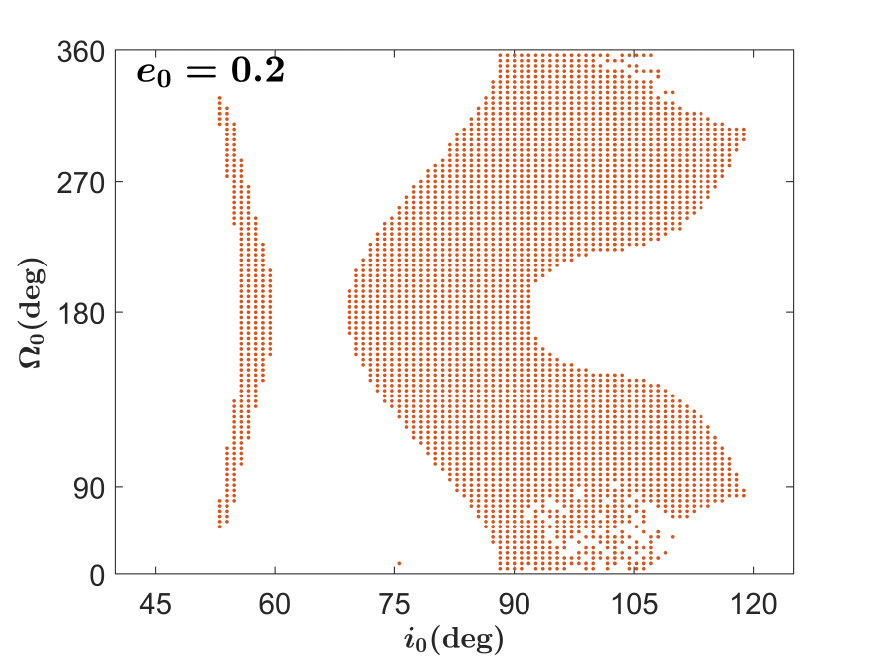}
    \includegraphics[width=5.6cm]{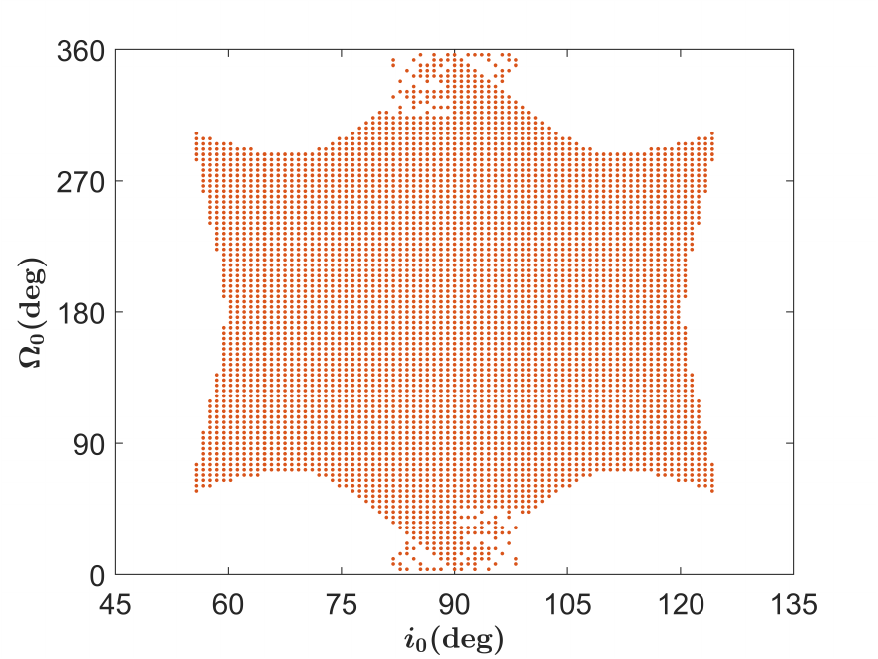}
    \includegraphics[width=5.6cm]{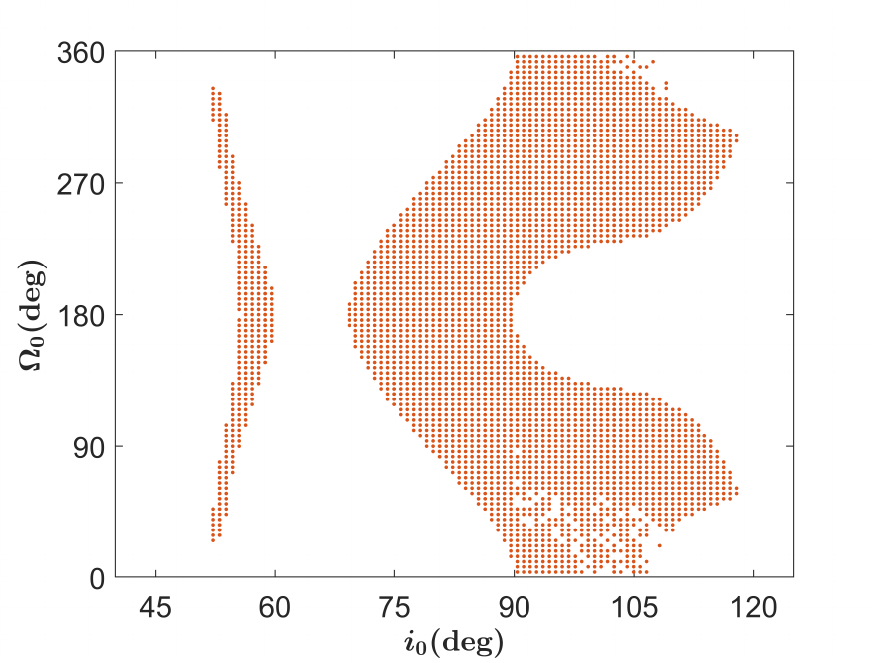}\\
    \includegraphics[width=5.6cm]{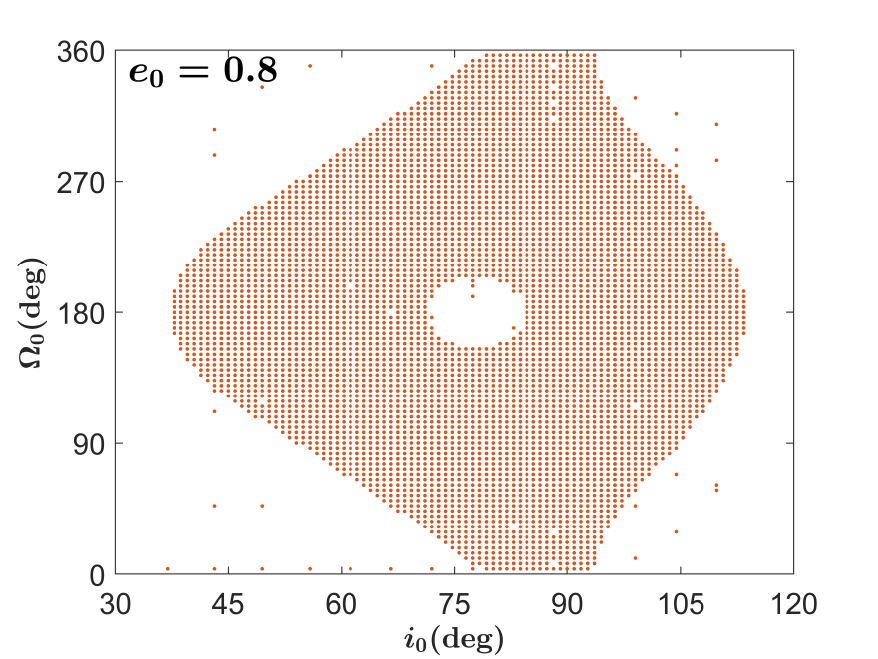}
    \includegraphics[width=5.6cm]{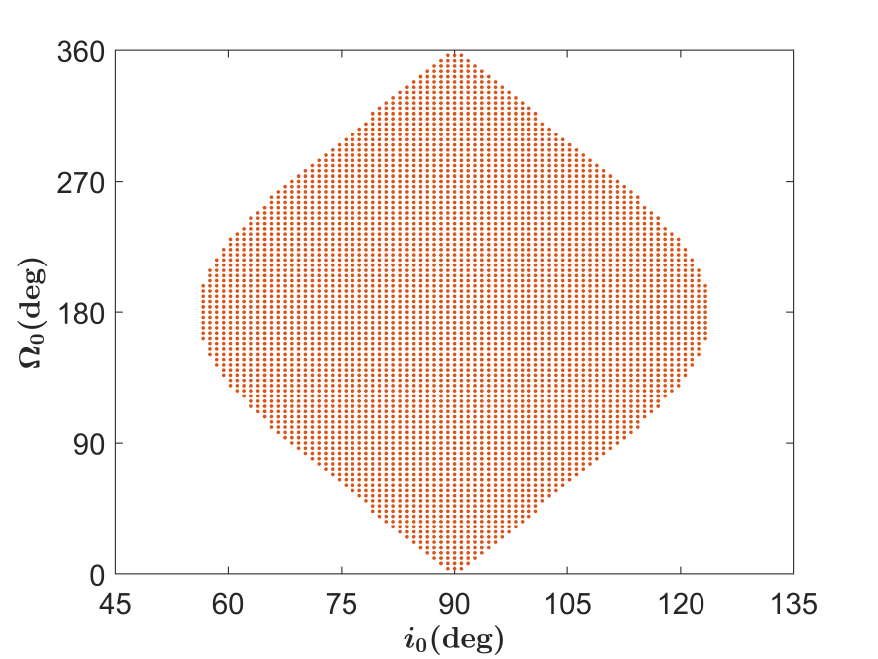}
    \includegraphics[width=5.6cm]{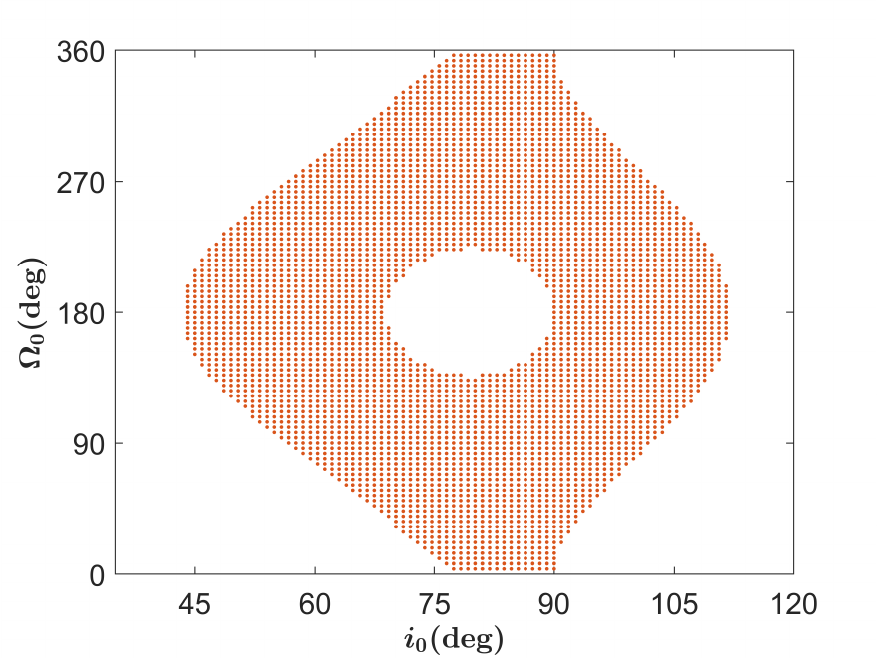}
    \caption{Orbital flipping regions distributed in the $(e_0,i_0)$ and $(i_0,\Omega_0)$ spaces. From the left to right, it corresponds to the results of N-body simulation, DA, and CDA, respectively. The model parameters are set as follows: $m_0=m_2=1.0\, m_{\odot},a_2=10\,\text{AU},e_2=0.2,i_2=\Omega_2=\omega_2=0^\circ,a_1=1.0\,\text{AU}$. The middle-row panels with $e_0=0.2$ stand for the case of orbit flipping in the low-eccentricity regime and the bottom-row panels with $e_0=0.8$ represent the case of orbit flipping in the high-eccentricity regime.
 }
\label{fig:2}
\end{figure*}

\section{High-order dynamical model}
\label{sec:3}

In this section, we aim to develop a high-order dynamical model up to dotriacontapole order in semimajor axis ratio with inclusion of Brown Hamiltonian corrections up to octupole-order coupling terms. This model can be considered as a realization of explicit form about the dynamical model developed in \citet{lei2018modified}. 

\subsection{Dynamical model with Brown corrections}
\label{sec:3-1}

To investigate the long-term dynamics of test particles, it is required to perform averaging over the periods of the inner and outer orbits. The first average is realized by
\begin{equation}
    \langle {\cal R} \rangle = \frac{1}{{2\pi }}\int_0^{2\pi } {{\cal R}\left( {1 - {e_1}\cos {E_1}} \right){\rm d}{E_1}}, 
        \label{eq:7}
\end{equation}
and the second average is performed through 
\begin{equation}
    \left\langle {\left\langle {\cal R} \right\rangle } \right\rangle  = \frac{1}{{2\pi }}\int_0^{2\pi } { {\left\langle {\cal R} \right\rangle } \frac{{{{\left( {1 - {{e_2}^2}} \right)}^{3/2}}}}{{{{\left( {1 + {e_2}\cos {f_2}} \right)}^2}}}{\rm d}{f_2}}. 
        \label{eq:8}
\end{equation}
The disturbing function after twice averaging without considering nonlinear effects associated with the inner and outer binaries determines the classical double-averaged model. Such a model works well in triple systems of high hierarchies. However, when the mass of the perturbing body becomes comparable to that of the central object, the hierarchy of system becomes mild. In this case, the averaging over the outer orbit removes significant perturbations experienced by the test particle, leading to significant discrepancy between the results of calssical double-averaged model and $N$-body simulations \citep{luo2016double,lei2018modified}. To address this problem, it is necessary to introduce an additional correction term \citep{soderhjelm1975three,krymolowski1999studies,luo2016double,breiter2015secular,cuk2004secular}, which is called Brown Hamiltonian \citep{tremaine2023}.

Brown Hamiltonian comes from the nonlinear effects of evection terms appearing in the single-averaged disturbing function \citep{tremaine2023}. To derive Brown Hamiltonian up to octupole-octupole coupling terms, we perform a Taylor expansion of the single-averaged disturbing function around the averaged orbit as follows:
\begin{equation}
    {\cal R}=\left\langle{{\cal R}}\right\rangle+\frac{{\partial \left\langle {\cal R} \right\rangle }}{{\partial {e_1}}}\delta {e_1}+\frac{{\partial \left\langle {\cal R} \right\rangle }}{{\partial {i_1}}}\delta {i_1}+\cdots
        \label{eq:9}
\end{equation}
where the second and third terms on the right-hand side are the nonlinear coupling effects caused by the periodic oscillations of eccentricity and inclination. According to mean element theory \citep{kozai1959motion}, the periodic oscillations including $\delta{e_1}$ and $\delta {i_1}$ can be obtained by
\begin{equation}\label{eq:11}
\begin{aligned}
    \delta {e_{1}} &= \int {\left( {\frac{{{\rm d}{e_1}}}{{{\rm d}{f_2}}} - {{\left\langle {\frac{{{\rm d}{e_1}}}{{{\rm d}{f_2}}}} \right\rangle }_{{f_2}}}} \right){\rm d}{f_2}},\\
    \delta {i_{1}} &= \int {\left( {\frac{{{\rm d}{i_1}}}{{{\rm d}{f_2}}} - {{\left\langle {\frac{{{\rm d}{i_1}}}{{{\rm d}{f_2}}}} \right\rangle }_{{f_2}}}} \right){\rm d}{f_2}},
\end{aligned}       
\end{equation}
where the subscript $f_2$ means that the average is performed over one period of the true anomaly. Based on Lagrange planetary equations \citep{brouwer1961methods,murray1999solar}, we can obtain  
\begin{equation}\label{eq:13}
\begin{aligned}
    \frac{{{\rm d}{e_1}}}{{{\rm d}{f_2}}} &=  -\frac{m_2}{m_0} \frac{\eta_1 a_1t_\text{ZLK}}{{{n_2a_2^3}{e_1}{{\left( {1 + {e_2}\cos {f_2}} \right)}^2}}}\frac{{\partial \left\langle {{{\cal R}}} \right\rangle }}{{\partial {\omega_1} }},\\
    \frac{{{\rm d}{i_1}}}{{{\rm d}{f_2}}} &= \frac{m_2}{m_0}\frac{{t_\text{ZLK}a_1}}{{{n_2}\eta_1a_2^3 {{\left( {1 + {e_2}\cos {f_2}} \right)}^2}}}\left[ {\cot {i_1}\frac{{\partial \left\langle {{{\cal R}}} \right\rangle }}{{\partial {\omega_1} }} - \csc {i_1}\frac{{\partial \left\langle {{{\cal R}}} \right\rangle }}{{\partial {\Omega_1} }}} \right],
\end{aligned} 
\end{equation}
where $\eta_1=\sqrt{1-e_1^2}$, ${n_1}$ and $n_2$ are the mean motion of the test particle and the perturber, respectively, and $t_\text{ZLK}$ is the time-scale of ZLK oscillations given by

\begin{equation}\label{eq:35}
    t_\text{ZLK}\sim\frac{1}{n_1}\left(\frac{m_0}{m_2}\right)\left(\frac{a_2}{a_1}\right)^3\left(1-e_2^2\right)^{3/2}
\end{equation}

By substituting the quadrupole- and octupole-order disturbing function into equation (\ref{eq:9}), we obtain three correction terms, called Brown Hamiltonian corrections: the quadrupole-quadrupole coupling correction, the quadrupole-octupole coupling correction, and the octupole-octupole coupling correction. Those high-order coupling corrections can be obtained in a similar manner. It should be mentioned that there is a gauge freedom in canonical transformations \citep{tremaine2023}, reflecting the dependence of the coupling correction on the choice of fictitious time. In particular, choice (3) in \citet{tremaine2023} leads to the simplest form of nonlinear coupling corrections. After considering the gauge freedom, we find that the quadrupole-octupole coupling correction term vanishes and the expressions of the quadrupole-quadrupole and octupole-octupole coupling terms can be largely simplified. 

In summary, the long-term Hamiltonian up to fifth order in semimajor axis ratio $\alpha$ with consideration of Brown Hamiltonian corrections up to the octupole-octupole coupling order can be written as
\begin{equation}
{\left\langle {\left\langle {\cal R} \right\rangle} \right\rangle } = {\cal C}_0 {\cal F},
\end{equation}
where ${\cal C}_0$ is a constant coefficient, given by
\begin{equation}\label{eq:14-2}
{\cal C}_0=\frac38\frac{{\cal G}m_2}{a_2}{\left( {\frac{{{a_1}}}{{{a_2}}}} \right)^2}\frac{1}{{{{\left( {1 - e_2^2} \right)}^{3/2}}}},
\end{equation}
which can be used to normalize the Hamiltonian, and the normalized Hamiltonian ${\cal F}$
is composed of six parts:
\begin{equation}
\begin{aligned}
    {\cal F} &= {\cal F}_\text{quad}+\varepsilon_\text{oct}{\cal F}_\text{oct}+\varepsilon_\text{hexa}{\cal F}_\text{hexa}+\varepsilon_\text{dotr}{\cal F}_\text{dotr} \\
    &+\varepsilon_\text{quad-quad}{\cal F}_\text{quad-quad}+\varepsilon_\text{oct-oct}{\cal F}_\text{oct-oct}.
    \label{eq:14-1}
\end{aligned}
\end{equation}
In particular, the quadrupole-order double-averaged Hamiltonian (corresponding to second order in $\alpha$) is
\begin{equation}\label{eq:15-1}
{\cal F}_{\rm quad}  =  {\frac{1}{6}\left( {2 + 3{e_1^2}} \right)\left( {3 \cos^2{i_1} - 1} \right) + \frac{5}{2}{e_1^2}{{\sin }^2}{i_1}\cos{2\omega _1}},
\end{equation}
the octupole-order double-averaged term (corresponding to third order in $\alpha$) is
\begin{equation}\label{eq:15-2}
\begin{aligned}
{\cal F}_{\rm oct} &= {\cal A}_{3,1} \left[ { {\cal B}_{3,1}\cos \left( {{\Omega _1} - {\omega _1}} \right)}  { +  {\cal B}_{3,2} \cos \left( {{\Omega _1} + {\omega _1}} \right)} \right]\\
&+{\cal A}_{3,2}\left[ { {\cal B}_{3,3} \cos \left( {{\Omega _1} - 3{\omega _1}} \right)}  { +  {\cal B}_{3,4} \cos \left( {{\Omega _1} + 3{\omega _1}} \right)} \right],
\end{aligned}
\end{equation}
the hexadecapole-order Hamiltonian (corresponding to fourth order in $\alpha$) is given by
\begin{equation}\label{eq:15-4}
\begin{aligned}
{\cal F}_{\rm hexa}&={\cal B}_{4,1}+ {\cal B}_{4,2}\cos 2{\Omega _1}+ {\cal B}_{4,3}\cos 2{\omega _1}+ {\cal B}_{4,4}\cos 4{\omega _1}\\
&+ {\cal A}_{4,1}\left[ {{\cal B}_{4,5}\cos \left( {2{\omega _1} + 2{\Omega _1}} \right)}  {+ {\cal B}_{4,6}\cos \left( {2{\omega _1} - 2{\Omega _1}} \right)} \right]\\
&+ {\cal A}_{4,2}\left[ {{\cal B}_{4,7}\cos \left( {4{\omega _1} + 2{\Omega _1}} \right)}   { + {\cal B}_{4,8}\cos \left( {4{\omega _1} - 2{\Omega _1}} \right)} \right],
\end{aligned}
\end{equation} 
and the dotriacontapole-order Hamiltonian (fifth order in $\alpha$) is given by
\begin{equation}\label{eq:15-6}
\begin{aligned}
{\cal F}_{\rm dotr}&={\cal A}_{5,1}\left[{\cal B}_{5,1} \cos{(\Omega_1+5\omega_1)}- {\cal B}_{5,2}\cos{\left(\Omega_1-5 \omega_1\right)}\right]\\
&+{\cal A}_{5,2}\left[ {\cal B}_{5,3} \cos {(3 \Omega_1+5 \omega_1)}- {\cal B}_{5,4} \cos{(3 \Omega_1-5 \omega_1)}\right]\\
&+{\cal A}_{5,3} \left[{\cal B}_{5,5} \cos{(3 \Omega_1-\omega_1)}-{\cal B}_{5,6} \cos{(3 \Omega_1+\omega_1)}\right]\\
&+{\cal A}_{5,4} \left[{\cal B}_{5,7} \cos{(\Omega_1-\omega_1)}- {\cal B}_{5,8} \cos{(\Omega_1+\omega_1)}\right]\\
&+{\cal A}_{5,5} \left[{\cal B}_{5,9} \cos{ (3\Omega_1-3\omega_1)}- {\cal B}_{5,10}\cos{(3\Omega_1+3\omega_1)}\right]\\
&+{\cal A}_{5,6}\left[ {\cal B}_{5,11} \cos{(\Omega_1+3 \omega_1)}-{\cal B}_{5,12} \cos{(\Omega_1-3 \omega_1)}\right].
\end{aligned}
\end{equation}
In addition, the nonlinear coupling terms (called Brown corrections) arised from quadrupole- and octupole-order evection terms are expressed as
\begin{equation}\label{eq:16-1}
\begin{aligned}
{\cal F}_\text{quad-quad} &= \frac{{3\eta_1 }}{{16}} \left( {1 + \frac{2}{3}{e_2^2}} \right) \cos{i_1}\\
&\times \left[ 2{\sin}^2{i_1}
+ {e_1^2}\left( {33 + 17\cos^2{i_1}}  +15{\sin}^2{i_1}\cos 2{\omega _1}\right)\right],
\end{aligned}
\end{equation}
and
\begin{equation}\label{eq:16-2}
\begin{aligned}
{\cal F}_\text{oct-oct}&=\frac{{25}}{{8192}} \eta_1 \cos{i_1} \left[ \frac{1}{4}\left({\cal Y}_1 - 10{\cos ^2}{i_1} {\cal Y}_2  { + 5{{\cos }^4}{i_1}{\cal Y}_3 }\right)\right.\\
& + \left. 25{e_1^2}{\sin}^2{i_1} {\cal Y}_4 \cos 2{\omega _1} { + \frac{{105}}{4}{\cal X}_7{e_1^4}{\sin}^4{i_1}  \cos 4{\omega _1}} \right],
\end{aligned}
\end{equation}
where $\eta_1 = \sqrt{1-e_1^2}$. In the above formula, the detailed expressions of ${\cal A}$, ${\cal B}$, ${\cal X}$ and ${\cal Y}$ are provided in Appendix \ref{appendix:A}. It should be pointed out that both the quadrupole-quadrupole and octupole-octupole coupling terms are axisymmetric (because both of them are independent on $\Omega_1$).

The significance of each part in equation (\ref{eq:14-1}) is controlled by a series of constant coefficients, given by
\begin{equation}\label{eq:15-3}
\begin{aligned}
{\varepsilon_\text{oct}} &= \left(\frac{{{a_1}}}{{{a_2}}}\right)\frac{{{e_2}}}{{1 - e_2^2}},\quad {\varepsilon _\text{hexa}} = \frac{3}{256}{\left( {\frac{{{a_1}}}{{{a_2}}}} \right)^2}\frac{1}{{{{\left( {1 - {e_2^2}} \right)}^2}}},\\
\varepsilon_\text{dotr}&=\frac{35}{16384}{\left( {\frac{{{a_1}}}{{{a_2}}}} \right)^3}\frac{e_2}{{{{\left( {1 - {e_2^2}} \right)}^3}}},\\
{\varepsilon_\text{quad-quad}}&=\left( {\frac{{{n_2}}}{{{n_1}}}} \right)\left( {\frac{{{m_2}}}{{{m_0} + {m_2}}}} \right)\frac{1}{{{{\left( {1 - e_2^2} \right)}^{3/2}}}},\\
{\varepsilon_\text{oct-oct}}&=\left( {\frac{{{n_2}}}{{{n_1}}}} \right)\left( {\frac{{{m_2}}}{{{m_0} + {m_2}}}} \right)\left(\frac{a_1}{a_2}\right)^2\frac{1}{{{{\left( {1 - e_2^2} \right)}^{7/2}}}},
\end{aligned}
\end{equation}
which indicates that the octupole- and dotriacontapole-order contributions may vanish when the perturber moves on a circular orbit. 

The quadrupole-order term ${\cal F}_{\rm quad}$ is the classical ZLK Hamiltonian \citep{Kozai1962}, which is used to describe the ZLK oscillations in highly hierarchical triple systems. The octupole-order Hamiltonian ${\cal F}_{\rm oct}$ is widely utilized to study the eccentric ZLK effects \citep{naoz2011hot,katz2011long,Lithwick&Naoz2011,naoz2013secular,li2014eccentricity,li2014chaos,naoz2016}. The quadrupole-quadrupole coupling term ${\cal F}_\text{quad-quad}$ corresponds to the classical Brown Hamiltonian correction \citep{brown1936c,luo2016double,cuk2004secular}, which has the same expression of equation (64) in \citet{tremaine2023}.

The averaged disturbing function given by equation (\ref{eq:14-1}) determines a two-degree-of-freedom dynamical model with $(g,G)$ and $(h,H)$ as two independent sets of canonical variables. Considering that the angular coordinate $l$ is a cyclic variable, its conjugate momentum $L$ becomes a motion integral, meaning that the semimajor axis $a_1$ of the test particle remains stationary during the long-term evolution. However, there is a unique conserved quantity, namely the Hamiltonian, making that the associated dynamical model is non-integrable. 

Without inclusion of Brown Hamiltonian corrections, it is possible to demonstrate that there is a symmetry inherent in the Hamiltonian model, given by \citep{sidorenko2018eccentric}
\begin{equation}\label{eq0}
{\cal F} (g,h,G,H) = {\cal F} (2\pi-g,h,G,-H) = {\cal F} (g,2\pi-h,G,-H),
\end{equation}
which ensures that the phase-space structures are symmetric with respect to $H=0$ (i.e., the line of $i_1=90^{\circ}$). However, the inclusion of Brown corrections (including ${\cal F}_\text{quad-quad}$ and ${\cal F}_\text{oct-oct}$) breaks the symmetry, showing that, under the complete Hamiltonian model developed in this study, the phase-space structures in the prograde and retrograde spaces are no longer symmetric.

We notice that \citet{will2017orbital,will2021higher} and \citet{conway2024higher} have derived long-term equations of motion (instead of Hamiltonian) for hierarchical triple systems up to dotriacontapole order in semimajor axis ratio and up to second order in mean motion ratio by means of two-scale method. Discussions about the equivalence among different models can be found in \citet{tremaine2023}.

\subsection{Validation of dynamical model}
\label{sec:3-2}

We compare $N$-body full model, the classical double-averaged model (without Brown corrections), and the corrected double-averaged model with inclusion of Brown corrections (hereinafter abbreviated as $N$-body, DA, and CDA, respectively). For all three models, we employed the RKF78 method with adaptive step-size control for practical numerical integration \citep{fehlberg1969classical}.

Figure \ref{fig:1} shows the time histories of the inclination $i_1$, eccentricity $e_1$ and the $z$-component of the orbital angular momentum $H$ of inner test particles under hierarchical triple systems with three levels of perturber's mass, representing dynamical systems from planetary to stellar perturbations. Please refer to the caption of Fig. \ref{fig:1} for the detailed setting of model parameters and initial conditions.

It is observed from Figure \ref{fig:1} that (a) in the case of planetary perturbation both the DA and CDA models can reproduce the results of $N$-body simulations (see the top panels with $m_2 = 0.001 m_\odot$), meaning that in the hierarchical planetary three-body problem the Brown corrections are negligible, and (b) as the perturber's mass increases from planetary to stellar scales, the evolutions of orbital elements under the CDA model are always in perfect agreement with the outcome of $N$-body simulations, while the results of DA model increasingly diverge from the $N$-body simulations (see the curves of $H$ shown in the middle- and bottom-row panels). In particular, when the mass of the perturber is comparable to that of the central object (see the bottom-row panels), the DA model predicts that the orbit may flip between prograde and retrograde status, which is totally different from the prediction of $N$-body simulation. The discrepancy between the DA and the $N$-body integration arises from the cumulative effect of those evection terms filtered out during the double-averaging process over long timescales \citep{luo2016double}. Because of $\varepsilon_\text{quad-quad} \propto \frac{m_2}{\sqrt{{m_0}\left({m_0+m_2}\right)}}$, the influence of Brown corrections increases with the mass of the perturbing body. In Figure \ref{fig:9}, we compare the modified models that incorporate different-level corrections. It shows that, compared to the original model, the corrected models exhibit better consistency with $N$-body simulation results.

Based on the discussions made above, the following conclusions can be summarized about dynamical models. When the mass of the perturbing body is relatively small compared to the central body, the Brown corrections have negligible effects, and the DA model can effectively describe the long-term behaviors of test particles. However, when the mass of the perturbing body is comparable to or greater than that of the central body, the DA model deviates significantly from the $N$-body integrations. By incorporating Brown corrections up to the octupole-octupole coupling terms, the resulting CDA model achieves nearly perfect alignment with $N$-body results. This demonstrates that the inclusion of quadrupole-quadrupole and octupole-octupole coupling corrections is essential for accurately modeling the secular evolution of test particles in mildly hierarchical triple systems.

\section{Eccentric ZLK effects}
\label{sec:4}

When the perturber moves on elliptic orbits and the disturbing function is truncated up to higher than octupole order in $\alpha$, the dynamical evolution of test particles undergoes periodic flips between prograde and retrograde spaces over long timescales, and in particular the eccentricity at the instant of flip can be extremely excited as high as close to 1. This intriguing phenomenon is known as the eccentric ZLK mechanism \citep{Lithwick&Naoz2011,katz2011long}. 

\subsection{Distribution of flipping orbits}
\label{sce:4.1}

It is known from the previous section that, the periodic terms removed during the second averaging process accumulate over long timescales, making the results of DA model deviate from the actual evolution. In general, the larger the mass of the perturbing body, the more pronounced the deviation. Under certain initial conditions, it can even alter the flip behavior of the orbit. Please see the bottom panels of Fig. \ref{fig:1}.

When the perturber's mass is comparable to that of the central object (corresponding to triple systems of mild hierarchies), Figure \ref{fig:2} shows the distribution of initial conditions of flipping orbits produced under the full $N$-body model (left-column panels), DA model (middle-column panels) and CDA model (right-column panels) in the $(e_0,i_0)$ and $(i_0,\Omega_0)$ spaces. Please refer to its caption for detailed setting of parameters.

It is observed from Figure \ref{fig:2} that the flipping regions under the DA model are symmetric about the line of ${i_0}=90^\circ$, but they do not agree with the results of $N$-body simulation. After incorporating the Brown corrections (CDA model, see the right-column panels), the flipping regions exhibit significant improvements and align almost perfectly with the $N$-body results. This indicates that the high-order averaged model with inclusion of Brown corrections developed in this work is accurate enough, and thus it offers a fundamental model to predict long-term dynamical behaviors of test particles under mildly hierarchical triple systems. 

Observing the flipping region produced under the $N$-body full model and the CDA model, we can see that (a) for all cases there is an asymmetrical distribution of flipping regions with respect to the line of $i_0 = 90^{\circ}$, (b) in the $(e_0,i_0)$ space there are low-eccenttricity, moderate-eccentricity and high-eccentricity flipping regimes, and in each regime some substructures can be observed, and (c) in the case of $e_0 = 0.2$ (low-eccentrcitiy configuration) there are two domains of flipping and, in the case of $e_0 = 0.8$, there is circular gap in the center of flipping region.

In the coming subsections, we will take advantage of the developed Hamiltonian model to explore the emergence of orbital flips based on dynamical system approach (i.e., Poincar\'e surfaces of section), and then utilize the method of perturbative treatment to study the asymmetrical distribution of flipping orbits caused by the eccentric ZLK effects under mildly hierarchical triple systems. Applications of perturbative treatment to understanding the eccentric ZLK effects under hierarchical planetary three-body systems can be found in \citet{sidorenko2018eccentric,lei2022,lei2022dynamical,lei2022quadrupole,huang2022orbital} and \citet{zhao2024zeipel}.

\begin{figure*}
    \includegraphics[width=7cm]{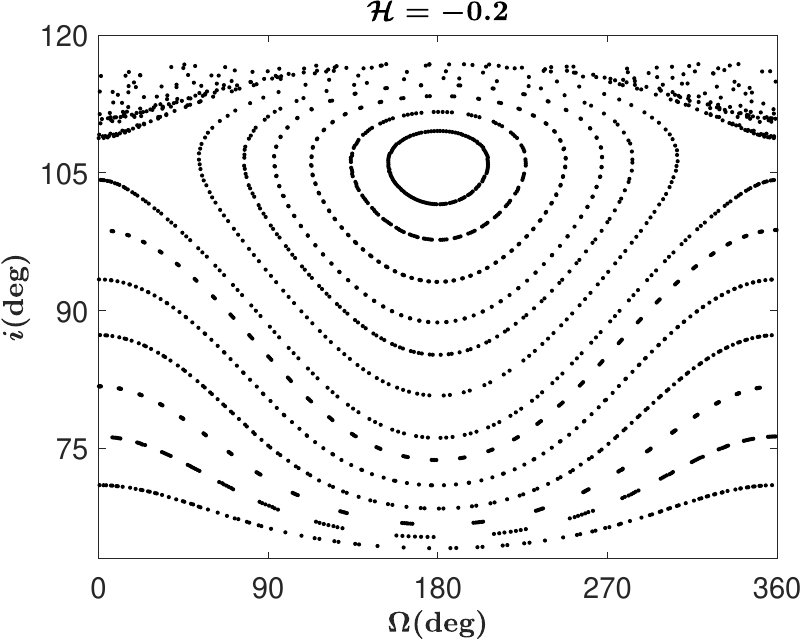}
    \includegraphics[width=7cm]{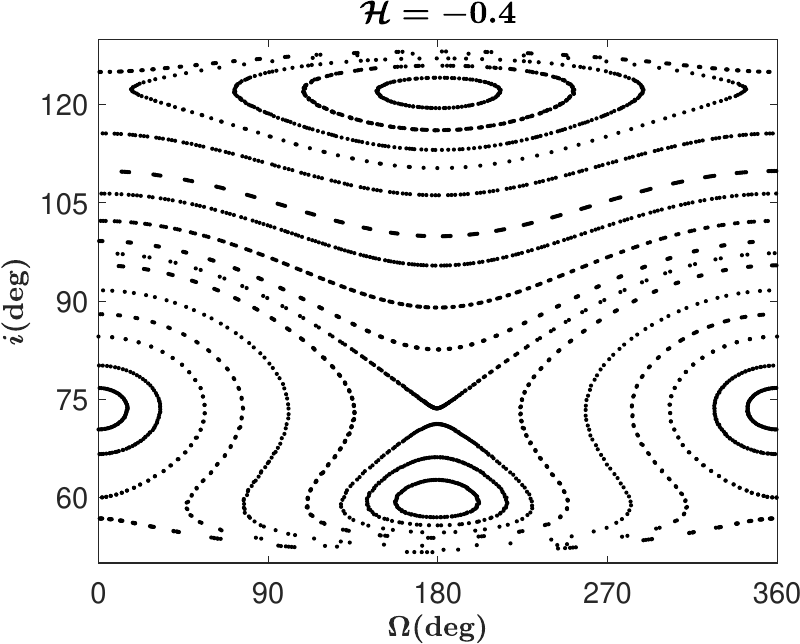}
    \\
    \includegraphics[width=7cm]{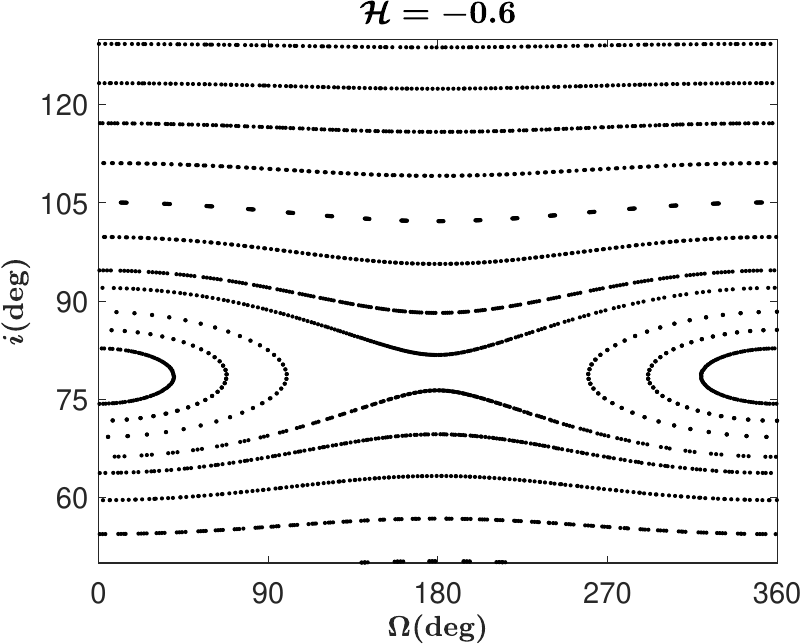}
    \includegraphics[width=7cm]{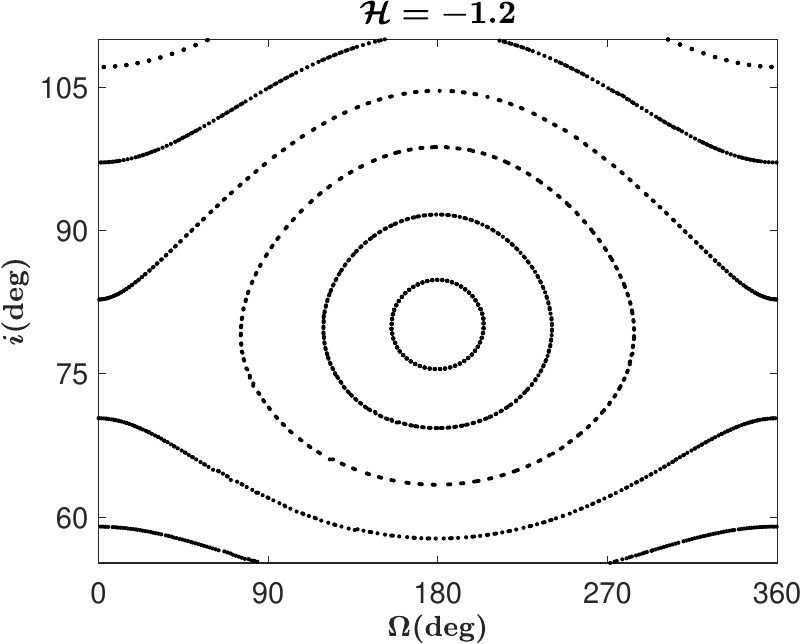}
    \caption{Poincaré sections shown in the $\left(\Omega, i\right)$ space for different levels of Hamiltonian. Those cycles crossing the lines of $i=90^{\circ}$ on the sections stand for flipping trajectories in the long-term evolution.}
    \label{fig:3}
\end{figure*}

\begin{figure}
    \includegraphics[width=8.5cm]{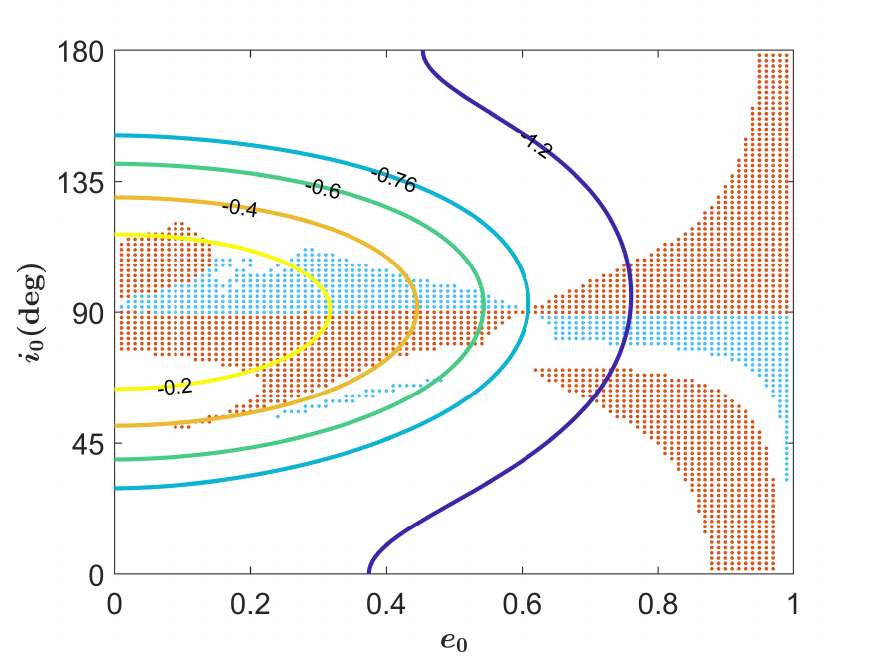}
    \caption{Orbital flipping region shown in the $\left(e_0,i_0\right)$ space obtained under the CDA model (corresponding to the upper-right panel of Figure \ref{fig:2}), together with representative level curves of Hamiltonian.}
    \label{fig:4}
\end{figure}

\subsection{Poincar\'e surfaces of section}
\label{sec:4.2}

The primary goal of this section is to investigate the origin of asymmetric structures exhibited by the flipping regions under the CDA model with massive perturber. This is achieved by probing the system's dynamical architecture through numerically constructed Poincaré sections, and establishing connections between their topological features and the morphological organization of flipping domains. 

According to Hamiltonian canonical relations, the equations of motion of test particles are given by \citep{morbidelli2002modern}
\begin{equation}\label{eq:22}
\frac{{\rm d}g}{{\rm d}t}=\frac{\partial{\cal H}}{\partial{G}},\quad \frac{{\rm d}G}{{\rm d}t}=-\frac{\partial{\cal H}}{\partial{g}},\quad \frac{{\rm d}h}{{\rm d}t}=\frac{\partial{\cal H}}{\partial{H}},\quad \frac{{\rm d}H}{{\rm d}t}=-\frac{\partial{\cal H}}{\partial{h}},
\end{equation}
which determines a two-degree-of-freedom dynamical system with ${\cal H}= - {\cal F}$. However, there is only one conserved quantity, namely the Hamiltonian. It means that the current dynamical system is non-integrable. For such a non-ntegrable dynamical system, the Poincar\'e surface of section serves as a powerful tool for numerically exploring the global structures within the phase space \citep{li2014chaos,lei2022}.

We define Poincar\'e surface of section by
\begin{equation}
    g=0,\quad \dot{g}>0.
    \label{eq:23}
\end{equation}
Fixing the Hamiltonian, we numerically integrate the equations of motion and record the intersection points on the $\left(h,H\right)$ plane to construct Poincar\'e sections. The resulting sections with Hamiltonian levels at ${\cal H}=-0.2,-0.4,-0.6,-1.2$ are shown in Figure \ref{fig:3}. It is observed that (a) dynamical structures shown in the $(\Omega,i)$ space are symmetric with respect to the line of $\Omega=180^{\circ}$; (b) dynamical structures are no longer symmetric with respect to the line of $i=90^{\circ}$, meaning that the structures in the prograde space are different from the ones in the retrograde space; (c) as the Hamiltonian decreases from ${\cal H} = -0.2$ to ${\cal H} = -1.2$, the number of libration islands changes from one to three and then return to one; (d) when the Hamiltoian is low the phase-space structure shown in the Poincar\'e section is similar to pendulum structure. In particular, those trajectories that cross the line of $i=90^{\circ}$ present flipping behaviors. Point (b) can help to understand the asymmetric distribution of flipping orbits observed in Fig. \ref{fig:2}, and point (d) shows that the eccentric ZLK effect in the high-eccentricity regime can be approximated by means of pendulum model\footnote{Actually, it is the region characterized by $H\ll1$. See Section \ref{sec:5}.} \citep{klein2024hierarchicala,klein2024hierarchicalb,basha2025kozai}.

Figure \ref{fig:4} illustrates the distribution of flliping orbits in the $(e_0,i_0)$ space, together with level curves of Hamiltonian used for producing Poincar\'e sections in Fig. \ref{fig:3}. By comparing Figures \ref{fig:3} and \ref{fig:4}, we can systematically elucidate the progressive evolution of flipping regions as a function of the Hamiltonian. Generally, with the Hamiltonian decreasing from ${\cal H}=-0.2$ to ${\cal H}=-1.2$, the Poincar\'e section undergoes sequential structural transitions from simplicity to complexity and back to simplicity. Initially, a resonant island resides in the upper portion of the section (in the retrograde space). Trajectories within this island, initiated at $\Omega=180^\circ$, generate two flipping domains, while trajectories in the exterior chaotic sea form irregular flipping regions near $\Omega=0^\circ$. As the Hamiltonian progressively decreases, the resonant island gradually shifts upward and diminishes in size, ceasing to induce orbital flips. Concurrently, two new resonant islands emerge in the lower part of section. Trajectories in the circulating regions flanking these upper islands develop two flipping regions initiated at $\Omega=180^\circ$, whereas the circulating region above the $\Omega=0^\circ$ island hosts an additional flipping region. As the Hamiltonian continues to decrease, the resonance islands centered at $\Omega=180^\circ$ disappear, leaving a single island of libration centered at $\Omega=0^\circ$. A subset of trajectories within this island crosses the line of $i=90^{\circ}$, leading to the emergence of a new flipping region initiated at $\Omega=0^\circ$. When the Hamiltonian is further reduced to about ${\cal H}=-0.76$, the resonance island vanishes entirely. At ${\cal H}=-1.2$, a new libration island appears at $\Omega=180^\circ$, within which part of trajectories undergo flipping. These trajectories, initiated at $\Omega=180^\circ$, generate two distinct flipping regions. Simultaneously, trajectories in the circulating regions develop an additional flipping region initiated at $\Omega=0^\circ$.

\begin{figure}
    \includegraphics[width=8cm]{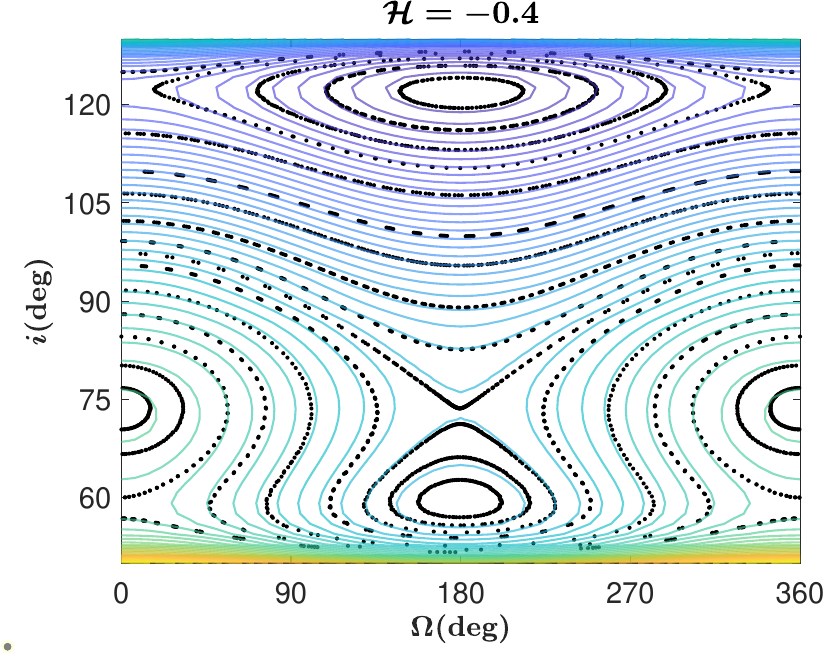}
    \caption{Comparison between Poincar\'e sections and contour plots of adiabatic invariant with the same Hamiltonian at ${\cal H}=-0.4$. The blue dots represent Poincaré sections (numerical results), while the colored solid lines correspond to contour lines of adiabatic invariant (analytical results).}
    \label{fig:5}
\end{figure}

\begin{figure*}
    \includegraphics[width=6.9cm]{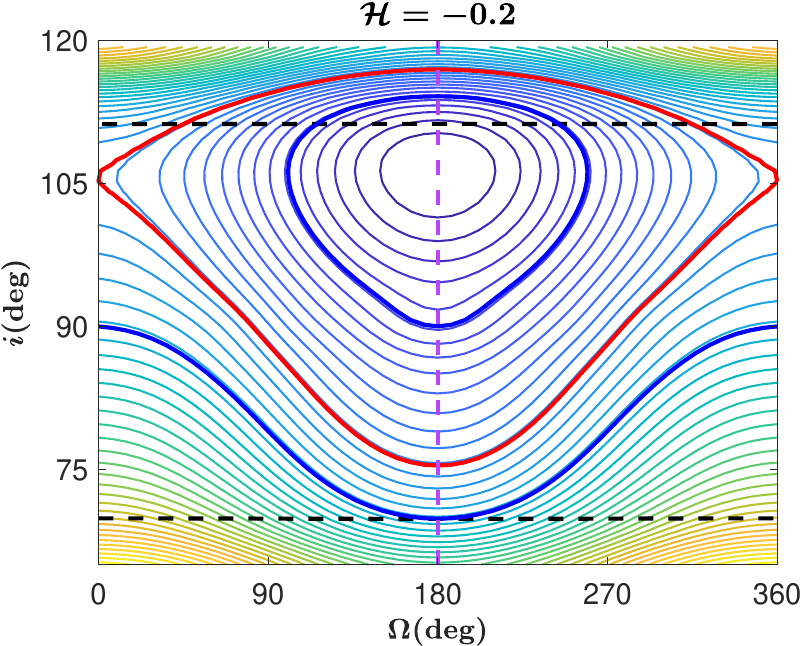}
    \includegraphics[width=7cm]{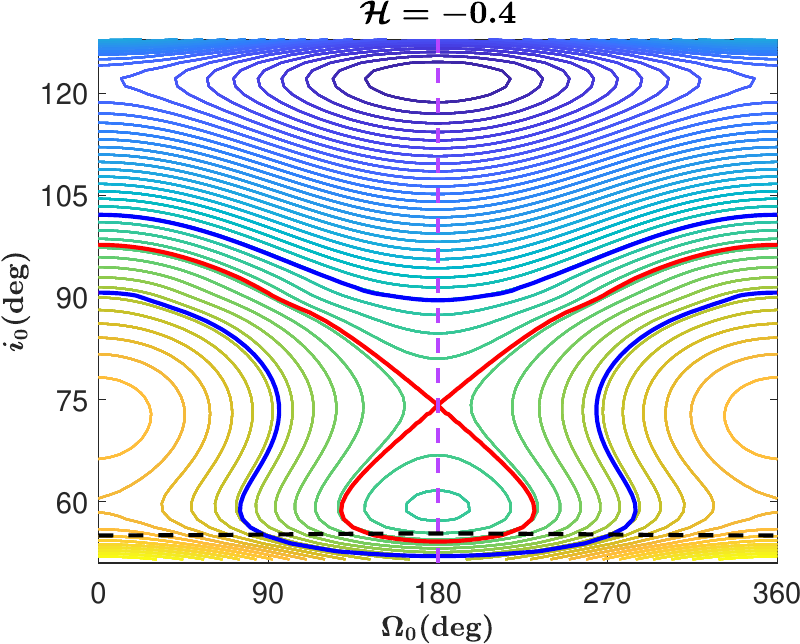}
    \\
    \includegraphics[width=7cm]{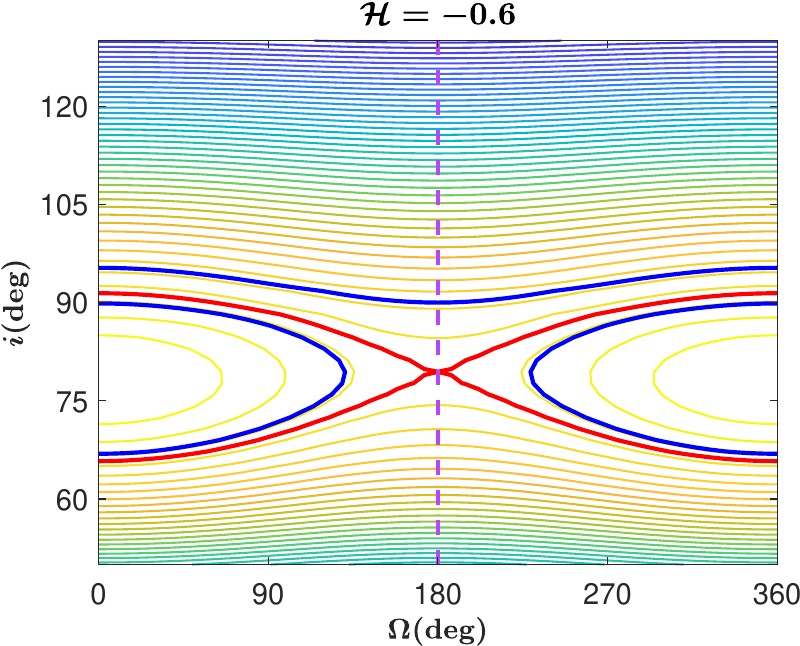}
    \includegraphics[width=7cm]{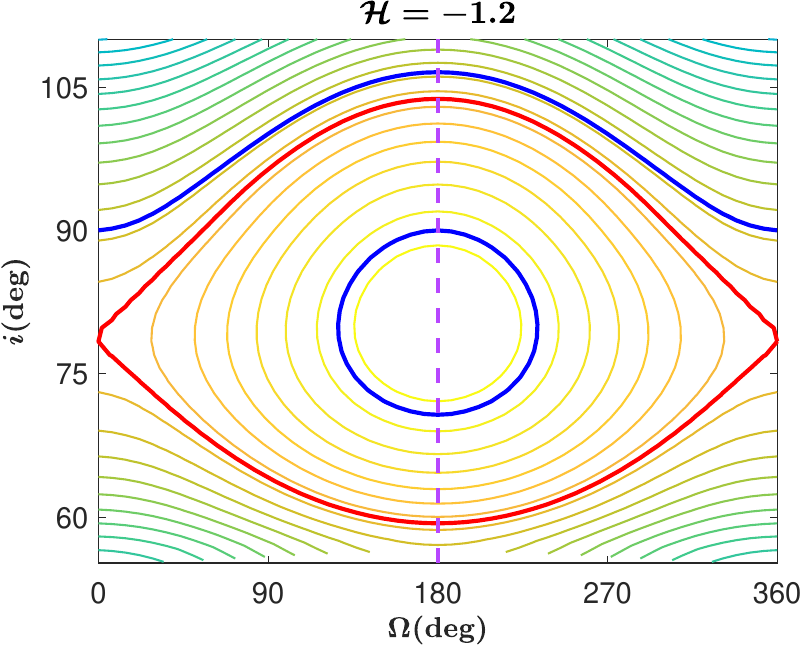}
    \caption{Phase portraits (level curves of adiabatic invariant) for four different levels of Hamiltonian at ${\cal H}=-0.2,-0.4,-0.6,-1.2$. The corresponding Poincar\'e sections can be found in Figure \ref{fig:3}. The red solid curves represent the dynamical separatrices between circulation regions and libration regions. The blue solid curves correspond to trajectories passing through the points $\left(0^\circ,90^\circ\right)$ in the phase space. The purple dashed lines mark the contour of $\left(180^\circ,90^\circ\right)$, while the black dashed curves indicate the isocontour of $G=\sqrt{1-e_0^2}$ within the phase portrait, where $e_0$ is taken as $0.2$ in this figure.}
    \label{fig:6}
\end{figure*}

\begin{figure*}
    \includegraphics[width=5cm]{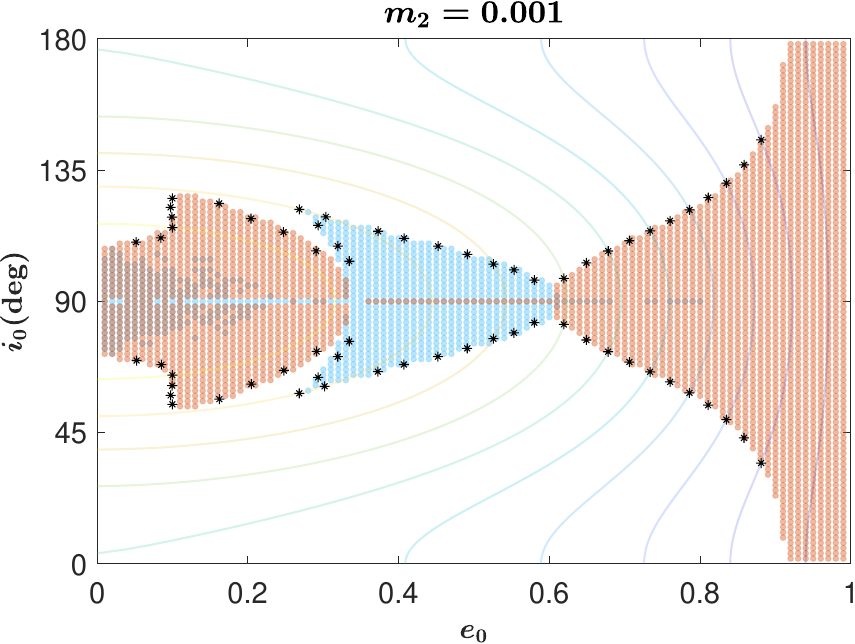}
    \includegraphics[width=5cm]{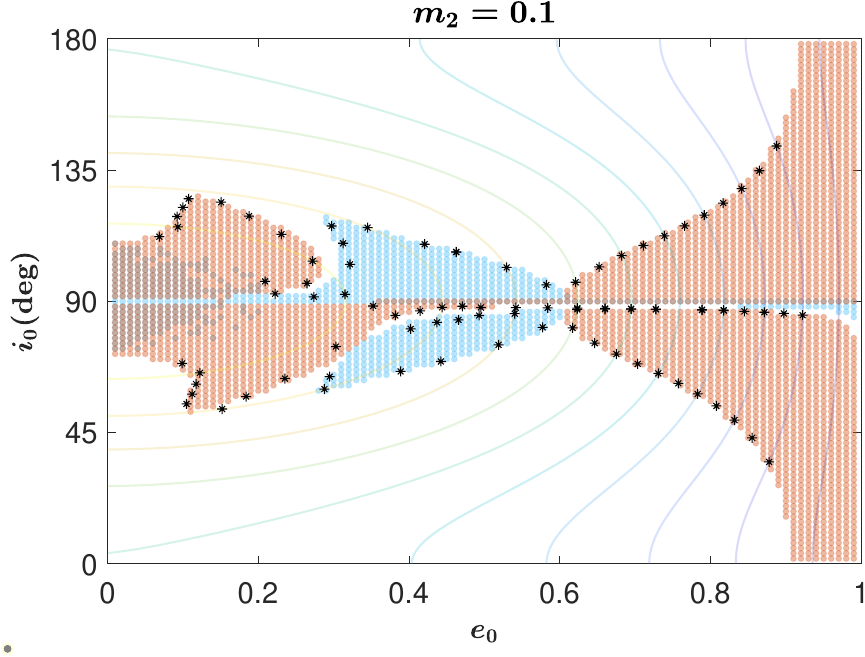}
    \includegraphics[width=5.05cm]{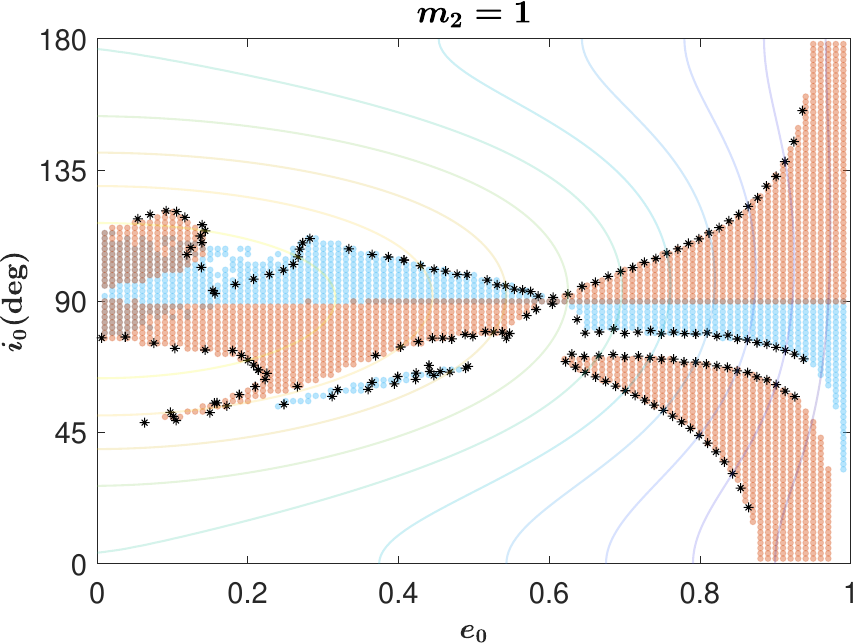}
    \\
    \hspace{0.1cm}
    \includegraphics[width=5.05cm]{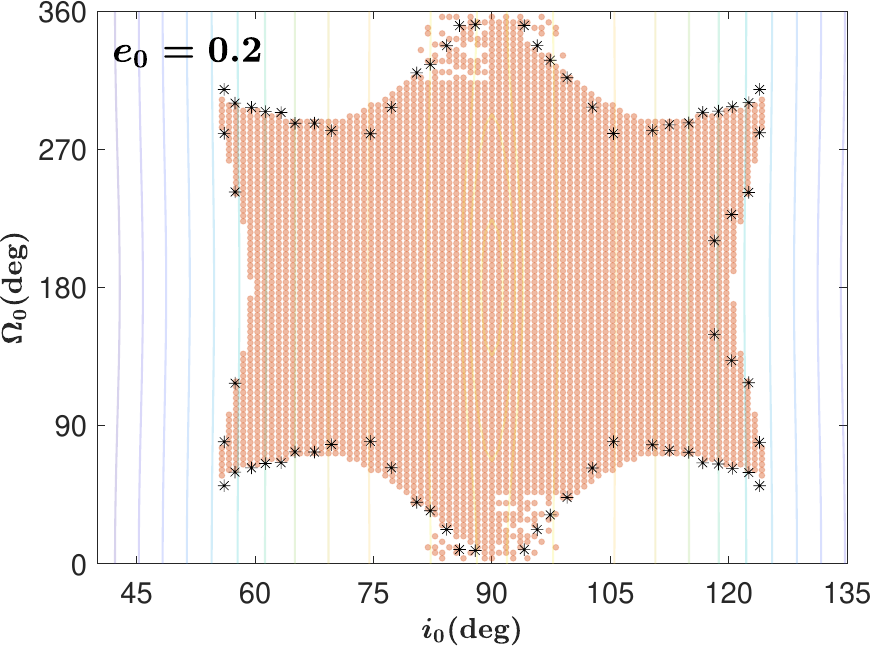}
    \includegraphics[width=5cm]{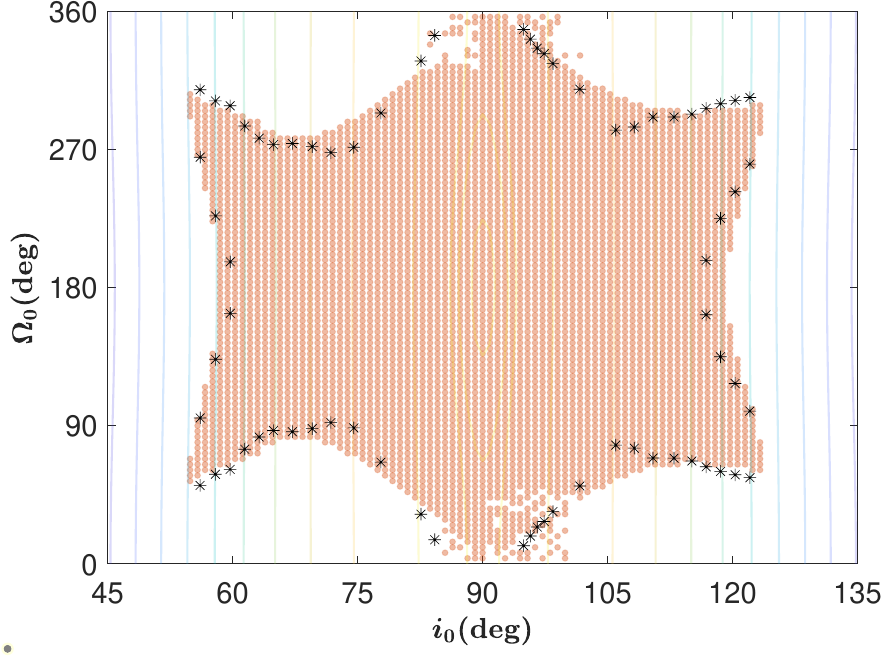}
    \includegraphics[width=5cm]{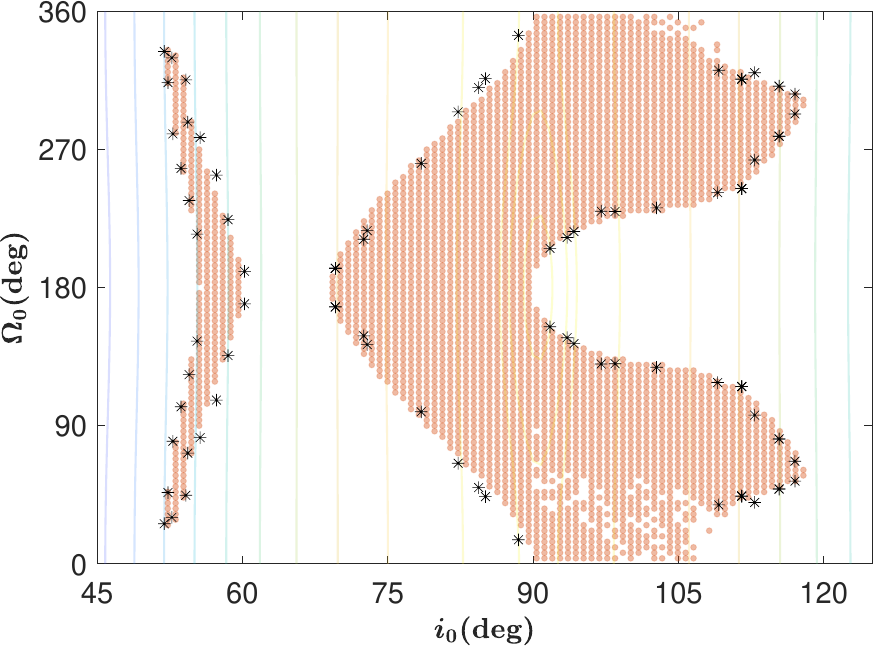}
    \\
    \hspace{0.1cm}
    \includegraphics[width=5.05cm]{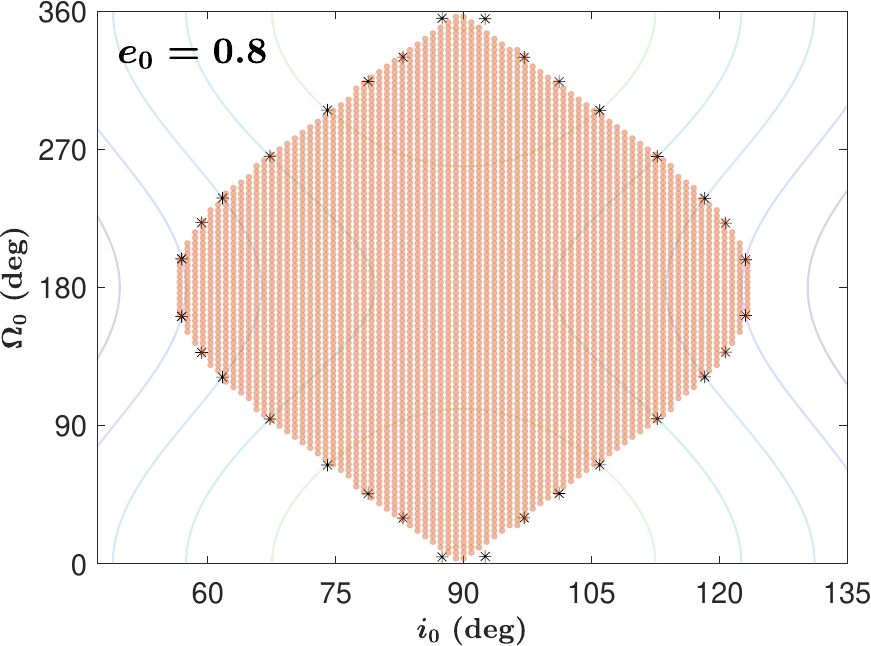}
    \includegraphics[width=5cm]{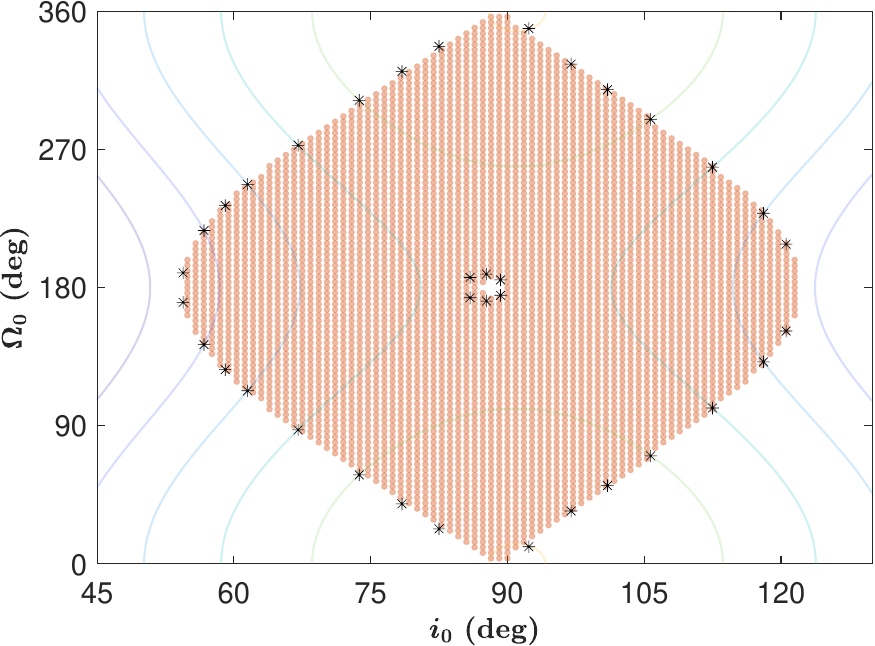}
    \includegraphics[width=5.05cm]{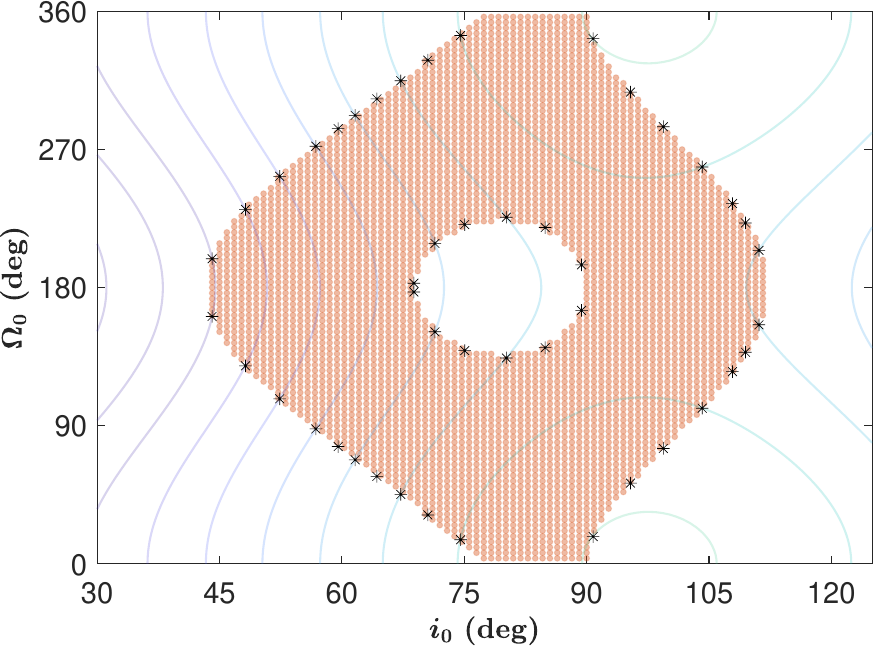}
    \caption{Orbital flipping regions derived from the CDA model and flip boundaries determined via analytical methods. From left to right, the perturber's mass changes from $m_2 = 0.001\,m_\odot$ to $m_2 = 1.0\,m_\odot$. The other parameters are consistent with the ones of Figure \ref{fig:2}. Black asterisks denote flip boundaries determined by analytical methods. Color-coded contours represent level curves of the Hamiltonian, shown as background.}
    \label{fig:7}
\end{figure*}

\begin{figure*}
    \includegraphics[width=5cm]{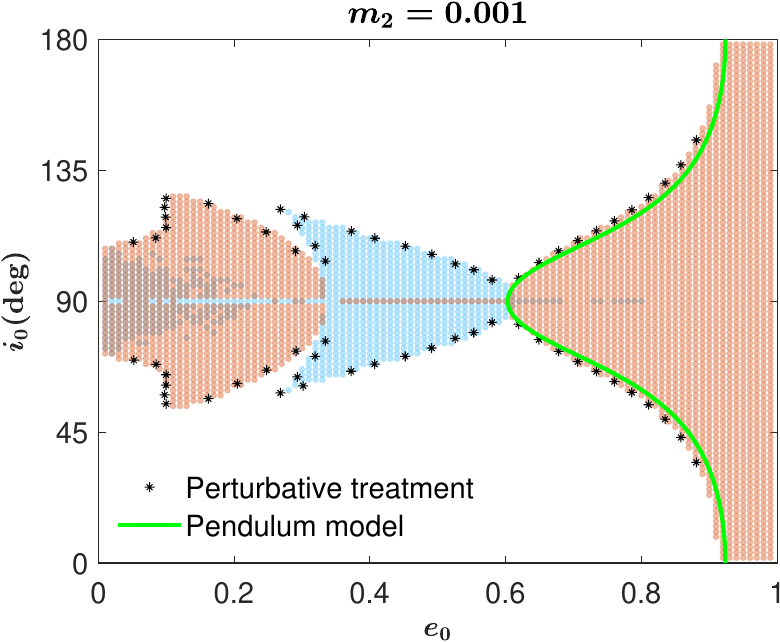}
    \includegraphics[width=5cm]{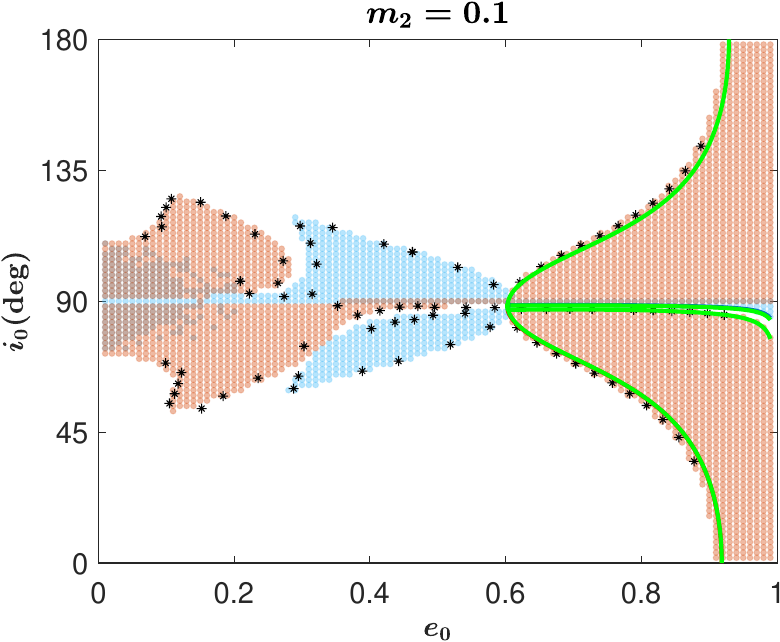}
    \includegraphics[width=5cm]{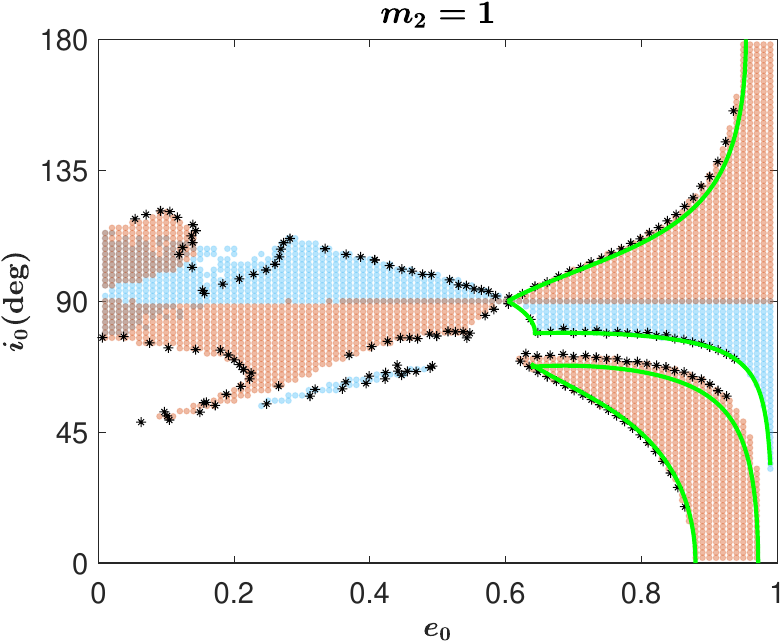}
    \\
    \hspace{0.2cm}\includegraphics[width=5cm]{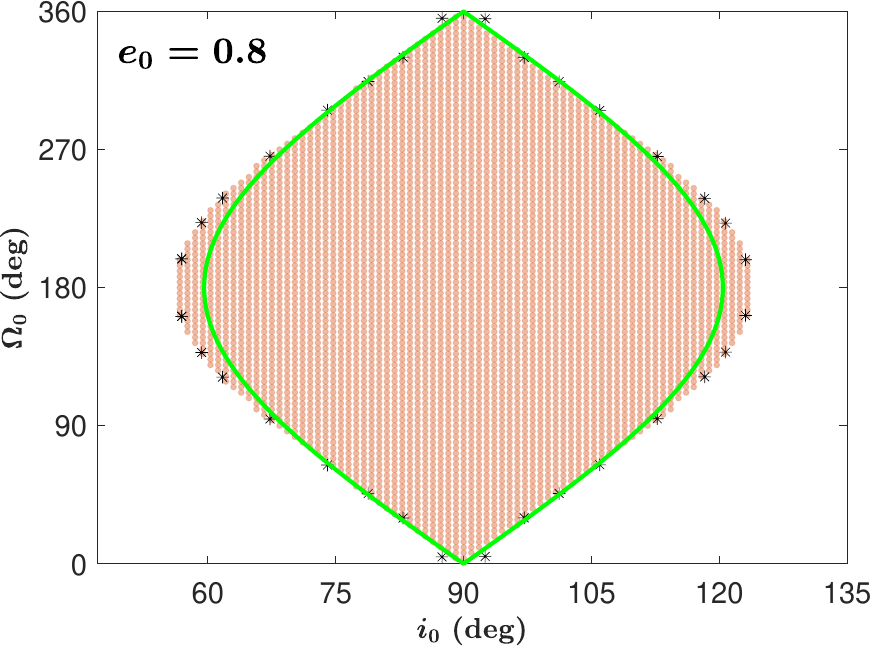}
    \hspace{-0.01cm}\includegraphics[width=4.9cm]{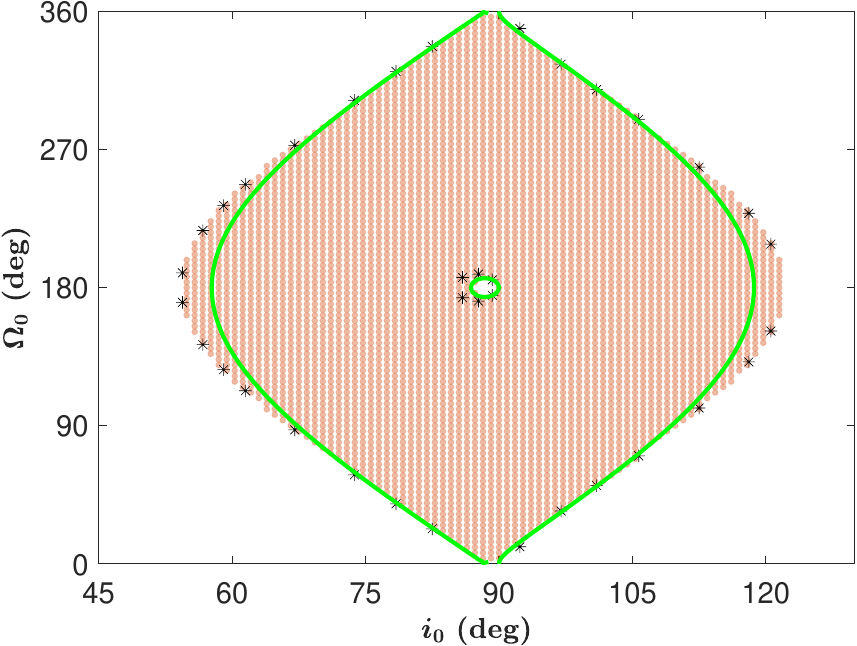}
    \hspace{0.15cm}\includegraphics[width=5cm]{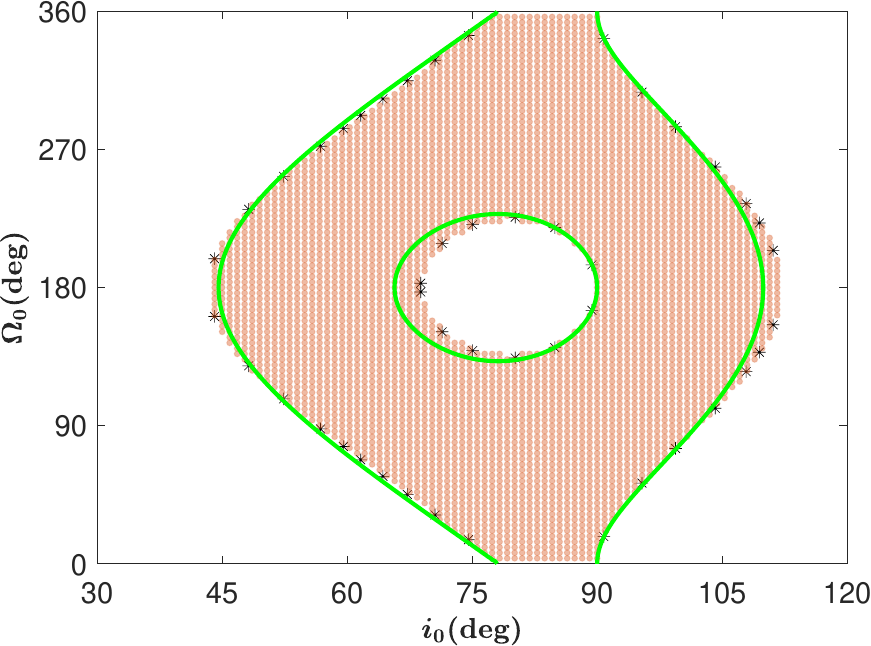}
    \caption{Comparison of flipping boundaries determined by the method of perturbative treatment and pendulum approximation, together with the numerical distribution of flipping orbits. The top-row panels show comparisons of flipping boundaries in the $\left(e_0, i_0\right)$ space and the lower-row panels show the comparsion in the $\left(i_0, \Omega_0\right)$ space for high-eccentricity configuration with $e_0 = 0.8$. In all panels, black asterisks denote the flipping boundaries determined by perturbative treatment and solid curves stand for the results of pendulum model.}
    \label{fig:8}
\end{figure*}

\subsection{Perturbative treatments}
\label{sec:4.3}

In this section, we employ the method of perturbative treatment to investigate the orbital flipping phenomenon of test particles and delineate the flipping boundaries within different parameter spaces. It is known that the orbital flipping phenomenon caused by the eccentric ZLK effect is induced by the octupole-order resonance, where the resonant angle $\sigma = h + \text{sign}\left(H\right)g$ librates around $0$ or $\pi$ \citep{sidorenko2018eccentric,lei2022,lei2022dynamical}. 

For convenience, the following set of variables is introduced,
\begin{equation}
    \begin{aligned}
        {\sigma_1}=&h+\text{sign}\left(H\right)g,&{\Sigma_1}=H
        \\
        {\sigma_2}=&g,&{\Sigma_2}=G-|H|
        \label{eq:24}
    \end{aligned}
\end{equation}
Under this set of variables, the Hamiltonian can be divided into two parts:
\begin{equation}
        \begin{aligned}
         {\cal H}\left(\sigma_1,\sigma_2,\Sigma_1,\Sigma_2\right) ={{\cal H}_\text{0}}\left(\sigma_2,\Sigma_1,\Sigma_2\right)+{\cal H}_\text{1}\left(\sigma_1,\sigma_2,\Sigma_1,\Sigma_2\right) 
         \label{eq:25}
    \end{aligned}
\end{equation}
where ${\cal H}_0$ denotes the kernel (or unperturbed) Hamiltonian, 
\begin{equation}
         {\cal H}_0 ={\cal H}_\text{quad}+{\cal H}_\text{quad-quad}+{\cal H}_\text{oct-oct}
         \label{eq:26}
\end{equation}
and ${\cal H}_1$ stands for the perturbation part,
\begin{equation}
         {\cal H}_1 ={\cal H}_\text{oct}+{\cal H}_\text{hexa}+{\cal H}_\text{dotr}.
         \label{eq:27}
\end{equation}
It is noted that the division of Hamiltonian is different from that in \citet{lei2022}. In this two-degree-of-freedom Hamiltonian system, it is observed that there exists a hierarchical timescale separation between the degrees of freedom: $(\sigma_1,\Sigma_1)$ corresponds to the slow degree of freedom, while $(\sigma_2,\Sigma_2)$ stands for the fast degree of freedom. In accordance with Wisdom's perturbation theory, slow variables can be treated as (constant) parameters within a period of fast degree of freedom \citep{wisdom1985perturbative,neishtadt1987change}. 
Regarding the fast degree of freedom $(\sigma_2,\Sigma_2)$, the Arnold action-angle variables can be defined by \citep{morbidelli2002modern}
\begin{equation}
    \Sigma_2^*=\frac{1}{2\pi}\int_{0}^{2\pi}\Sigma_2\text{d}\sigma_2,\quad \sigma_2^*=\frac{2\pi}{T}t
    \label{eq:28}
\end{equation}
where $T$ is the period of $\sigma_2$. Under the new set of action-angle variables, the Hamiltonian becomes
\begin{equation}
    {\cal H}\left(\sigma_1,\Sigma_1,\Sigma_2\right) ={{\cal H}_\text{0}}\left(\Sigma_1,\Sigma_2^*\right)+{\cal H}_\text{1}\left(\sigma_1,\Sigma_1,\Sigma_2^*\right),
    \label{eq:29}
\end{equation}
which is independent on $\sigma_2^*$, indicating that the action $\Sigma_2^*$ becomes an integral of motion, corresponding to an adiabatic invariant during the long-term evolution, denoted by \citep{henrard1993adiabatic}
\begin{equation}
    {\cal S}\left({\cal H},\sigma_1,\Sigma_1\right)=\int_{0}^{2\pi}\Sigma_2\left({\cal H},\sigma_1,\Sigma_1,\sigma_2\right)\text{d}\sigma_2
    \label{eq:30}
\end{equation}
The geometric meaning of the adiabatic invariant is the area enclosed by the curve $\Sigma_2\left(\sigma_2\right)$ within one period \citep{morbidelli2002modern}. Specifically, when the variables $\Sigma_1$, $\sigma_1$ and ${\cal H}$ are provided, the adiabatic invariant ${\cal S}$ can be produced by numerically integrating $\Sigma_2\left(\sigma_2\right)$ over one period of $\sigma_2$ from $0$ to $2\pi$ \citep{lei2022}.

Upon introducing the adiabatic invariant, this two-degree-of-freedom system exhibits two conserved quantities: ${\cal H}$ and ${\cal S}$. When either one of these quantities is specified by the system's initial conditions, the dynamical evolution is constrained to proceed along trajectories where the other quantity remains invariant. Consequently, the system's phase portrait can be obtained by plotting level curves of adiabatic invariant in the phase space $\left(\sigma_1,\Sigma_1\right)$ under a specified Hamiltonian \citep{wisdom1985perturbative,henrard1989motion,henrard1990semi}. 

In Figure \ref{fig:5}, phase portrait with the Hamiltonian at ${\cal H} = -0.4$ is presented together with its corresponding Poincar\'e section. It is observed that there is a good correspondence between the phase portrait and the associated Poincar\'e section, demonstrating the validity of Wisdom's perturbation theory in dealing with this dynamical system. As a result, the dynamical behaviours caused by eccentric ZLK effects can be analytically explored by analyzing phase portraits.

In particular, through the phase portrait, we can analyze the boundaries of orbital flipping regions. Figure \ref{fig:6} displays the phase portraits for Hamiltonian levels at ${\cal H}=-0.2,-0.4,-0.6,-1.2$. In these portraits, the variables $\left(\sigma_1,\Sigma_1\right)$ have been transformed into $\left(\Omega,i\right)$ coordinates. A comparative analysis with Figure \ref{fig:4} reveals that, as the Hamiltonian ${\cal H}$ decreases, the evolution of resonance islands in the phase portraits aligns precisely with the dynamical trends observed in the corresponding Poincaré sections. In Figure \ref{fig:6}, the red solid curves represent dynamical boundaries acting as the separatrix between resonance zones and libration regions, the blue solid curves depict trajectories passing through the points $\left(0^\circ, 90^\circ\right)$ and $\left(180^\circ, 90^\circ\right)$, the purple dashed lines correspond to $\Omega=180^\circ$, and the black dashed curves mark the $G=\sqrt{1-e_0^2}$ contour in the phase portrait. The intersection points between the purple dashed lines and solid curves determine the $i$ coordinates of the flipping boundaries in $(e_0, i_0)$ parameter space for initial conditions with $\Omega_0 = 180^\circ$. Given $i_0$ and the Hamiltonian ${\cal H}$, $e_0$ can be solved via Equation (\ref{eq:14-1}) to characterize the flipping boundaries. Meanwhile, the intersections of black dashed lines with solid curves define the flipping boundaries in the $\left(i, \Omega\right)$ space.

In Figure \ref{fig:7}, we present analytical boundaries by analyzing phase portraits (black stars) together with numerical ditribution of flipping orbits in the $(e_0,i_0)$ and $(i_0,\Omega_0)$ spaces under different levels of perturber's mass. For the case of $(i_0,\Omega_0)$ space, the initial eccentricity is fixed at $e_0 = 0.2$ standing for low-eccentricity configurations and at $e_0 = 0.8$ standing for high-eccentricity configurations. It is observed that, for all the considered cases covering from planetary to stellar perturbations, there is an excellent agreement between the analytical boundaries and the associated numerical distributions of flipping orbits. As the perturber's mass grows, the symmetry of flipping regions with respect to the line of $i_0 = 90^{\circ}$ gradually breaks down. 

\section{Pendulum approximation in the high-eccentricity regime}
\label{sec:5}

The eccentric ZLK effects in the high-eccentricity regime have been successfully approximated by a simple pendulum model under the octupole-level Hamiltonian model without Brown correction \citep{klein2024hierarchicala} and with consideration of quadrupole-quadrupole coupling term, namely the so-called Brown Hamiltonian correction \citep{klein2024hierarchicalb}.

In this section, we aim to extend the pendulum approximation made in \citet{klein2024hierarchicala,klein2024hierarchicalb} to the octupole-level Hamiltonian\footnote{Numerical simulations show that the hexadecapole- and dotriacontapole-level Hamiltonian have slight influence upon the distribution of flipping regions, thus they are not considered in the pendulum model for simplicity.} with consideration of quadrupole-quadrupole and octupole-octupole coupling terms in order to analyze the flipping regions in the high-eccentrcity regime. As a result, it becomes possible for us to compare the pendulum-approximation results with the analytical results of perturbative treatments discussed in the previous section.

In the limit of $j_z \ll 1$ (here $j_z$ is equal to $H$, namely the $z$-component of orbital angular momentum) and under the approximation of ${\cal R}_\text{quad}={\rm const}$, it is possible to perform averaging over the period of ZLK cycles at ${j_z}=0$ (pendulum approximation) for the equations of motion including quadrupole-quadrupole, octupole-octupole coupling corrections, leading to the averaged equations of motion,
\begin{equation}
\begin{aligned}
    {\dot{\Omega}_e} = &\langle{f_\Omega}\rangle j_z -\varepsilon_\text{quad-quad}\langle{f_\text{quad-quad}}\rangle -\varepsilon_\text{oct-oct}\langle{f_\text{oct-oct}}\rangle,\\
    {\dot{j_z}} = &-\varepsilon_\text{oct}\langle{f_j}\rangle \sin{\Omega_e},
    \label{eq:31}
\end{aligned}
\end{equation}
where $\langle{f_\Omega}\rangle$, $\langle{f_\text{quad-quad}}\rangle$, $\langle{f_\text{oct-oct}}\rangle$ and $\langle{f_j}\rangle$ are quantities related to the eccentricity vector $\bmath{e}$ and the dimensionless orbital angular momentum vector $\bmath{j}$ of the test particle (see Appendix \ref{appendix:B} for their expressions). The vectors $\bmath{e}$ and $\bmath{j}$ are given by \citep{klein2024hierarchicalb}
\begin{equation}\label{eq:32}
\begin{aligned}
{\bmath{e}} = & e_1\left(\cos{\Omega_e}\cos{i_e},\sin{\Omega_e}\cos{i_e},\sin{i_e}\right),\\
{\bmath{j}} = & \eta_1\left(\sin{i_1}\sin{\Omega_1},-\sin{i_1}\cos{\Omega_1}, \cos{i_1} \right),
\end{aligned}
\end{equation}
where $\eta_1=\sqrt{1-e_1^2}$. Note that, in all the calculations presented in this work, the initial value of $\omega_1$ is consistently assumed as zero. Consequently, $\Omega_e$ in the pendulum model is equivalent to $\Omega_1$.

From the structure of the equations, it is evident that this system reduces to a simple pendulum model. When the corrections are neglected (i.e., $\varepsilon_\text{quad-quad}=0,\varepsilon_\text{oct-oct}=0$), the resonance center of the simple pendulum is located at $j_z=0$. In the phase portrait, all trajectories crossing $j_z=0$ exhibit symmetry about $j_z=0$. This symmetry corresponds to the phenomenon observed in the DA model, where the flipping regions in both $\left(e_0, i_0\right)$ and $\left(i_0, \Omega_0\right)$ parameter spaces are symmetric about $i_0=90^\circ$. When Brown Hamiltonian corrections are considered, the resonance center of the pendulum model shifts according to the first equation. This shift breaks the symmetry of the phase portrait about $j_z=0$ (i.e. $i_0=90^\circ$), leading to asymmetric structures of flipping regions.

In Figure \ref{fig:8}, we make a comparison between simple pendulum model and the corrected Hamiltonian model under different conditions. The solid lines represent the flipping boundaries derived from the simple pendulum model, while the black asterisks denote the flipping boundaries derived from perturbative treatments discussed in the previous section.

It is observed from Figure \ref{fig:8} that the simple pendulum model can accurately reproduce the boundaries of orbital flipping in high-eccentricity regimes, demonstrating the validity of the simple pendulum approximation in these regions. This feasibility is further corroborated by Figures \ref{fig:4} and \ref{fig:6}, where the Poincar\'e sections and phase portraits of the system in high-eccentricity flipping regions (e.g., ${\cal H}=-1.2$) hold similar phase-space structures to a simple pendulum. Approximating such regimes with a pendulum model represents a rational and elegant simplification, greatly simplifying the model. 

It is noted that, under the simple pendulum model, the term involving $j_z$ in the following equation was truncated to the lowest-order term under the limit of $j_z \ll 1$ \citep{katz2011long}
\begin{equation}
    \dot{j_z}=\frac{75}{64}\varepsilon_\text{oct}\left[2j_yj_ze_z-e_y\left(\frac15-\frac85e_1^2+7e_z^2-j_z^2\right)\right]
    \label{eq:34}
\end{equation}
When the condition of $j_z\ll1$ is violated, equation (\ref{eq:34}) would introduce coupling terms between $\bmath{j}$ and $\bmath{e}$, thereby invalidating the simple pendulum approximation. For example, the condition of $j_z\ll1$ is not satisfied in the regions of low and moderate eccentricity, where the pendulum approximation cannot work. However, the analytical results derived from perturbative treatment can agree well with numerical distribution of flipping regions in the entire eccentricity space, as shown in Figure \ref{fig:8}. 

\section{Conclusions}
\label{sec:6}

In this study, we developed a long-term dynamical model up to the dotriacontapole order in terms of the semimajor axis ratio, incorporating Brown Hamiltonian corrections that include both quadrupole–quadrupole and octupole–octupole coupling terms. The resulting Hamiltonian is expressed in closed form with respect to the eccentricities of the inner and outer binaries. Within the framework of this Hamiltonian model, we systematically investigated orbit-flipping behavior induced by the eccentric ZLK mechanism, utilizing Poincar\'e sections, perturbative treatments, and pendulum approximation.

When the mass of the perturbing body is significantly smaller than that of the central body (corresponding to highly hierarchical triple systems), the influence of Brown corrections is negligible, and thus the double-averaged (DA) approximation reliably captures the long-term dynamical evolution of test particles. However, as the perturbing body’s mass increases to become comparable to or greater than that of the central body (corresponding to mildly hierarchical systems), the DA model shows substantial discrepancies relative to direct $N$-body simulations. In such cases, inclusion of Brown corrections up to the octupole–octupole coupling terms yields the corrected double-averaged (CDA) model, which markedly improves the accuracy of long-term predictions, exhibiting excellent agreement with $N$-body results.

Based on the formulated Hamiltonian model, orbital flips are investigated for different levels of perturber's mass. The results show that incorporating Brown Hamiltonian corrections significantly refines the structure of flipping regions. As the systems range from planetary-scale to stellar-scale perturbations, flipping regions evolves from a symmetric structure with respect to the $i=90^{\circ}$ line to a distinctly asymmetric one. These structures are analyzed using both numerical technique (Poincar\'e sections), and analytical method (perturbative treatments). Notably, there is a perfect correspondence between the numerical structures shown in Poincar\'e sections and the analytical structures arising in phase portraits, indicating that the eccentric ZLK effects in mildly hierarchical triple systems can be understood within an analytical framework. A comparison among different levels of perturbations reveals that the Brown Hamiltonian corrections are primarily responsible for breaking the symmetry of flipping regions. 

In the high-eccentricity regime, the pendulum approximation is extended to the formulated Hamiltonian model, showing an excellent agreement among the numerical distribution of flipping orbits, analytical boundaries derived from perturbative treatments, and analytical results of pendulum model.

\section*{Acknowledgements}
We would like to thank Prof. Scott Tremaine for suggesting us to make a comparison between our model with those existing nonlinear models, Dr. Ygal Y. Klein for helpful suggestions about the pendulum model, Dr. Boaz Katz and the anonymous reviewer for their helpful comments that significantly improve the quality of this work. This work is financially supported by the National Natural Science Foundation of China (Nos. 12233003 and 12073011) and the China Manned Space Program with grant no. CMS-CSST-2025-A16.
\section*{Data Availability}
The codes used in this article could be shared on reasonable request.
\bibliographystyle{mnras}
\bibliography{reference} 
\appendix
\section{Hamiltonian model}
\label{appendix:A}
\begin{equation*}
\begin{aligned}
&{\cal A}_{3,1}=\frac{5}{{64}} {e_1} \left( {{3}{e_1^2} + 4} \right),\quad {\cal A}_{3,2}=- \frac{{175}}{{64}}{e_1^3}\\
&{\cal A}_{4,1}=35{e_1^2}{e_2^2}\left( {{e_1^2} + 2} \right),\quad {\cal A}_{4,2}=\frac{{735}}{4}{e_1^4}{e_2^2}\\
&{\cal A}_{5,1}=-1386 e_1^5 \left(3 e_2^2+4\right),\quad {\cal A}_{5,2} = 693 e_1^5 e_2^2\\\
&{\cal A}_{5,3}=-14 e_1 e_2^2 \left(5 e_1^4+20 e_1^2+8\right)\\
&{\cal A}_{5,4}=4 e_1 \left(5 e_1^4+20 e_1^2+8\right) \left(3 e_2^2+4\right)\\
&{\cal A}_{5,5}=7 e_1^3 e_2^2 \left(3 e_1^2+8\right),\quad {\cal A}_{5,6}=42 e_1^3 \left(3 e_1^2+8\right) \left(3 e_2^2+4\right)
\end{aligned}
\end{equation*}
\begin{equation*}
\begin{aligned}
&{\cal B}_{3,1}=1 - 11c - 5c^2 + 15c^3,\quad {\cal B}_{3,2}=1 + 11c - 5c^2 - 15c^3\\
&{\cal B}_{3,3}=1 - c - c^2 + c^3,\quad {\cal B}_{3,4}=1 + c - c^2 - c^3\\
&{\cal B}_{4,1}={\left( {2 + 10{e_1^2} + \frac{15}{4} {e_1^4}} \right)\left( {2 + 3{e_2^2}} \right)} \left( 3 - 30c^2 + 35c^4 \right)\\
&{\cal B}_{4,2}=-5 {e_2^2}\left( {4 + 20{e_1}^2 + \frac{15}{2} {e_1^4}} \right) \left( 1 - 8c^2 + 7c^4 \right)\\
&{\cal B}_{4,3}=-35{e_1^2}\left( {2 + {e_1^2}} \right)\left( {2 + 3{e_2^2}} \right) \left( 1 - 8c^2 + 7c^4 \right)\\
&{\cal B}_{4,4}=\frac{{735}}{4}{e_1^4} {s^4} \left( {2 + 3{e_2^2}} \right)\\
&{\cal B}_{4,5}={{\left( {1 + c } \right)}^2}\left( {1 - 7c + 7c^2} \right),\quad {\cal B}_{4,6}={{\left( {1 - c} \right)}^2}\left( {1 + 7c + 7c^2} \right)\\
&{\cal B}_{4,7}={s^2}{{\left( {1 + c} \right)}^2},\quad {\cal B}_{4,8}={s^2} {{\left( {1 - c} \right)}^2}
\end{aligned}
\end{equation*}
\begin{equation*}
\begin{aligned}
&{\cal B}_{5,1}={s^4} (1+c),\quad {\cal B}_{5,2}=- {s^4} (1 - c)\\
&{\cal B}_{5,3}=-{s^2}(1+c)^3,\quad {\cal B}_{5,4}={s^2} (1-c)^3\\
\end{aligned}
\end{equation*}
\begin{equation*}
\begin{aligned}
&{\cal B}_{5,5}=(1-c)^2 \left(15 c^3+21 c^2+5 c-1\right)\\
&{\cal B}_{5,6}=(1+c)^2 \left(15 c^3-21 c^2+5 c+1\right)\\
&{\cal B}_{5,7}=105 c^5-21 c^4-126 c^3+14 c^2+29 c-1\\
&{\cal B}_{5,8}=105 c^5+21 c^4-126 c^3-14 c^2+29 c+1\\
&{\cal B}_{5,9}=-(1-c)^3 \left(45 c^2+54 c+13\right)\\
&{\cal B}_{5,10}=(1+c)^3 \left(45 c^2-54 c+13\right)\\
&{\cal B}_{5,11}=(1+c)^2 \left(15 c^3-21 c^2+5 c+1\right)\\
&{\cal B}_{5,12}=(1-c)^2 \left(15 c^3+21 c^2+5 c-1\right)\\
\end{aligned}
\end{equation*}
\begin{equation*}
\begin{aligned}
&{\cal Y}_1={8 {\cal X}_1  + 8{e_1^2}{\cal X}_2 + {e_1^4} {\cal X}_3 },\quad{\cal Y}_2={\cal X}_4 + 8{e_1^2}{\cal X}_5  + {e_1^4} {\cal X}_6\\
&{\cal Y}_3={\cal X}_7\left( {8 + 8{e_1^2} + 47{e_1^4}} \right),\quad {\cal Y}_4={ - 9{e_1^2} {\cal X}_8  - 2 {\cal X}_9 }  + 3 c^2 {\cal X}_7\left( {2 + 5{e_1^2}} \right)
\end{aligned}
\end{equation*}
where
\begin{equation*}
\begin{aligned}
&{\cal X}_1=232 + 523{e_2^2} + 75{e_2^4},\quad {\cal X}_2=296 + 1319{e_2^2} + 195{e_2^4}\\
&{\cal X}_3=68632 + 125173{e_2^2} + 17685{e_2^4}\\
&{\cal X}_4=832 + 1528{e_2^2} + 216{e_2^4},\quad {\cal X}_5=72 + 213{e_2^2} + 31{e_2^4}\\
&{\cal X}_6=7384 + 7261{e_2^2} + 957{e_2^4},\quad {\cal X}_7=200 + 335{e_2^2} + 47{e_2^4}\\
&{\cal X}_8=168 + 147{e_2^2} + 19{e_2^4},\quad {\cal X}_9=392 + 203{e_2^2} + 23{e_2^4}
\end{aligned}
\end{equation*}
and
\begin{equation*}
c=\cos{i_1},\quad s=\sin{i_1}.
\end{equation*}

\section{Pendulum model}
\label{appendix:B}
\begin{equation*}
\begin{aligned}
&\langle{f_j\rangle}=\frac{15\pi}{128\sqrt{10}}\frac{\left(4-11C_K\right)\sqrt{6+4C_K}}{K(x)},\quad \langle{f_\Omega\rangle}=\frac{6E(x)-3K(x)}{4K(x)}\\
&\langle{f_\text{quad-quad}\rangle}=\frac{27}{64} \left(1+\frac{2}{3}e_2^2\right) \left(1+8C_K+\langle{e_z^2\rangle}\right)
\end{aligned}
\end{equation*}
\begin{alignat*}{1}
    \langle{f_\text{oct-oct}\rangle}=&\frac{75}{{1048576}} \left\{ 20 \langle{e_z^2\rangle} \left[(11685 C_K+688) e_2^4\right. \right.\\
    &\left.+(81157 C_K+4464) e_2^2+8 (4211 C_K-48)\right]\\
    &+ 20 \left[C_K (2853 C_K+128)+120\right] e_2^4\\
    &+ 4\left[C_K (98713 C_K+2432)+4184\right] e_2^2\\
    &+ 32 \left[C_K (4799 C_K-1664)+232\right]\\
    &\left.+5 \left(61317 e_2^4+430565 e_2^2+212120\right) \langle{e_z^4\rangle}\right\}
\end{alignat*}
where
\begin{equation*}
\begin{aligned}
\langle{e_z^2\rangle}=&\frac{2}{15} \left[\frac{(3+2 C_K) E(x)}{K(x)}-5 C_K\right]\\
\langle{e_z^4\rangle}=&\frac{4}{675} \left[5 C_K (13 C_K-3)-\frac{2 (16 C_K^2+18 C_K-9) E(x)}{K(x)}\right]
\end{aligned}
\end{equation*}
with
\begin{equation*}
x=\frac{3-3 C_K}{3+2 C_K},\quad C_K=e_1^2\left(1-\frac{5}{2}\sin^2{i_1}\sin^2{\omega_1}\right).
\end{equation*}
Here, $K(x)$ and $E(x)$ are the first and second types of complete elliptic integrals, respectively.

\bsp	
\label{lastpage}
\end{document}